\documentclass[%
 reprint,
 amsmath,amssymb,
pra,
]{revtex4-2}

\usepackage{graphicx}
\usepackage{dcolumn}
\usepackage{bm}

\usepackage{physics}
\usepackage{xcolor}
\usepackage{multirow}
\usepackage{esvect}
\usepackage{longtable}
\usepackage{hyperref}
\renewcommand{\vv}[1]{\boldsymbol{\mathbf{#1}}}

\begin{document}

\preprint{APS/123-QED}

\title{Molecular enhancement factors for  P, T-violating eEDM in BaCH$_3$ and YbCH$_3$ symmetric top molecules}

\author{Yuly Chamorro}
\affiliation{Van Swinderen Institute for Particle Physics and Gravity, University of Groningen, 9747 AG, Groningen, The Netherlands}
\author{Anastasia Borschevsky}
\affiliation{Van Swinderen Institute for Particle Physics and Gravity, University of Groningen, 9747 AG, Groningen, The Netherlands}
\author{Ephraim Eliav}
\affiliation{School of Chemistry, Tel Aviv University, 69978 Tel Aviv, Israel}
\author{Steven Hoekstra}
\affiliation{Van Swinderen Institute for Particle Physics and Gravity, University of Groningen, 9747 AG, Groningen, The Netherlands}
\author{Nicholas R. Hutzler}
\affiliation{Division of Physics, Mathematics, and Astronomy, California Institute of Technology, Pasadena, California 91125, USA}
\author{Lukáš F. Pašteka}
\affiliation{Department of Physical and Theoretical Chemistry, Faculty of Natural Sciences, Comenius University, Mlynská dolina, 84215
Bratislava, Slovakia}
\email{lukas.f.pasteka@uniba.sk}

\date{\today}

\begin{abstract}

High-precision tests of fundamental symmetries are looking for the parity- (P), time-reversal- (T) violating electric dipole moment of the electron (eEDM) as proof of physics beyond the Standard Model. Particularly, in polyatomic molecules, the complex vibrational and rotational structure gives the possibility to reach high enhancement of the P, T-odd effects in moderate electric fields. Additionally, it is possible to increase the statistical sensitivity by using laser cooling. In this work, we calculate the P, T-odd electronic structure parameters $W_\mathrm{d}$ and $W_\mathrm{s}$ for the promising candidates BaCH$_3$ and YbCH$_3$ for the interpretation of future experiments. We employ high-accuracy relativistic coupled cluster methods and systematically evaluate the uncertainties of our computational approach. 
Compared to other Ba- and Yb-containing molecules, BaCH$_3$ and YbCH$_3$ exhibit larger $W_\mathrm{d}$ and $W_\mathrm{s}$ associated to increased covalent character of the M--C bond. 
The calculated values are $3.22\pm 0.11 \times 10^{24}\frac{h\text{Hz}}{e\text{cm}}$ and $13.80\pm 0.35 \times 10^{24}\frac{h\text{Hz}}{e\text{cm}}$ for $W_\mathrm{d}$, and $8.42\pm0.29$~$h$kHz and $45.35\pm1.15$~$h$kHz for $W_\mathrm{s}$, in  BaCH$_3$ and YbCH$_3$, respectively. The robust, accurate, and cost-effective computational scheme reported in this work makes our results suitable for extracting the relevant fundamental properties from future measurements and also can be used to explore other polyatomic molecules sensitive to various violations of fundamental symmetries.
\end{abstract}

\maketitle

\section{Introduction}

The Standard Model of particle physics (SM) is known to be the most successful theory in describing the universe at the smallest scale. This model predicts all the known fundamental particles and explains their interactions via three of the four fundamental forces. However, despite its successful descriptions, SM is known to be an incomplete theory. Several well established experimentally observed facts, such as the matter-antimatter asymmetry and the existence of dark matter \cite{bertone2010particle} and dark energy \cite{sola2013cosmological} are not described by the SM. In particular, the matter-antimatter asymmetry requires an amount of charge-parity (CP) violation incompatible with the SM \cite{dine2003origin}. The incompleteness of the SM is both an incentive and an opportunity to look for physics beyond the Standard Model (BSM), also known as new physics.

High-precision tests of fundamental symmetries are a very effective means of probing BSM physics \cite{ginges2004violations}. Specifically, precision experiments in the sub-microhertz level using atoms and molecules are searching for the electric dipole moment of the electron (eEDM). The eEDM violates both time-reversal symmetry (T) and parity symmetry (P) and, assuming the CPT theorem, the eEDM thus also violates CP symmetry. In the SM, the eEDM is highly suppressed, and the predicted value is far too small to be measured using current experimental techniques. This has been estimated to be in the order of magnitude of $|d_{\text{e}}^{\text{SM}}| <  10^{-40}$ \cite{yamaguchi2020large}. On the other hand, the BSM theories predict values in the experimental reach \cite{commins1999electric,bernreuther1991electric}, and a measurement of a non-zero value would be an incontrovertible proof of new physics \cite{pospelov2014ckm}. 

Presence of the eEDM induces an EDM on paramagetic molecules \cite{sachs1959implications,salpeter1958some} which is enhanced due to the internal electric fields. It has been previously shown that this enhancement grows with the atomic number as $Z^3$ \cite{bouchiat1974parity} making systems containing heavy atoms ideal for measuring the eEDM. Additionally, the experimental signal is also enhanced when using close-lying opposite parity states -- more information can be found in the work of Sandars \cite{sandars1965electric,sandars1968electric}. In atoms, opposite-parity electronic states are split by $\sim 2$ eV, while, in molecules, opposite-parity rotational states are split by $\sim 10^{-5}$ eV, typically. Consequently, some molecular states can offer a dramatically higher enhancement than the enhancement found in atoms \cite{sandars1967measurability,demille2015diatomic}. Motivated by such enhancements, numerous experiments are being developed using molecules containing heavy atoms. The first experiments using diatomic molecules were performed on TlF in Oxford \cite{sandars1967measurability,harrison1969experimental,hinds1976limits,hinds1980experiment}, and the current most stringent result has been set in ThO by ACME collaboration; $|d_{\text{e}}^{\text{ThO}}| < 1.1 \times 10^{-29}$ $\textit{e}\cdot$cm \cite{andreev2018improved}. Other investigated diatomic molecules include YbF \cite{kara2012measurement}, HfF$^+$ \cite{cairncross2017precision}, ThO \cite{andreev2018improved}, RaF \cite{garcia2020spectroscopy,isaev2010laser}, and BaF \cite{aggarwal2018measuring}.

Some diatomic molecules, such as PbO \cite{bickman2009preparation}, ThO \cite{andreev2018improved}, and HfF$^+$ \cite{cairncross2017precision}, have almost degenerate opposite parity (excited) eigenstates, called $\Omega$-doublets, which are used to measure the eEDM. These parity doublets have a small splitting which makes it possible to fully mix (or polarize) them in moderate electric fields. Additionally, the use of $\Omega$-doublets give the possibility to cancel many systematic effects. Unfortunately, due to the complex structure of the $\Omega$-doublets states, it is not possible to take advantage of laser cooling to improve the experimental sensitivity. On the other hand, measurements of the eEDM in diatomic molecules in their ground state, such as BaF \cite{aggarwal2018measuring}, YbF \cite{hudson2011improved,kara2012measurement}, and RaF \cite{isaev2010laser}, cannot make use of the $\Omega$-doublet states.

Polyatomic molecules emerge as good candidates for eEDM experiments \cite{kozyryev2017precision,hutzler2020polyatomic}. In contrast to diatomic molecules, in polyatomic molecules it is possible to measure the eEDM in the long-lived close-lying opposite parity eigenstates ($K$-doublets). In addition, previous works \cite{ellis2001main,isaev2017laser,isaev2016polyatomic,kozyryev2016proposal,mitra2020direct,augenbraun2021observation} have discussed the feasibility of laser cooling of metal-containing molecules composed of a single metal atom bound to a single molecular ligand, such as the symmetric top molecules BaCH$_3$ and YbCH$_3$. This is also supported by that fact the lighter CaCH$_3$ and MgCH$_3$ exhibit quasi-diagonal Frank--Condon factors \cite{isaev2016polyatomic}. 

The enhancement of the eEDM interaction in atoms and molecules is usually expressed in terms of the electronic structure parameter $W_\mathrm{d}$. The scalar-pseudoscalar (S-PS) interaction between the electrons and the nucleons is also a P,T-violating interaction expressed in terms of the parameter $W_\mathrm{s}$ \cite{kozlov1995parity} and is similarly enhanced in atoms and molecules. Both parameters are needed to connect the measured energy shift due to P,T-violation to the eEDM value. Since
$W_\mathrm{d}$ and $W_\mathrm{s}$ can not be measured experimentally, it is necessary to use electronic structure methods to calculate them.  
 
In this work, we report the $W_\mathrm{d}$ and $W_\mathrm{s}$ parameters for the promising eEDM candidates for future experiments, the BaCH$_3$ and YbCH$_3$ molecules. Since the enhancement factor depends mainly on the identity of the heavy atom, it is expected that BaCH$_3$ and YbCH$_3$ have similar enhancement factors comparable to other isolectronic molecules relevant for eEDM experiments, such as BaF, BaOH, and YbF, YbOH \cite{kozlov1994enhancement,haase2021systematic,denis2019enhancement,gaul2020ab}, respectively. However, unlike in the diatomic molecules BaF and YbF, the symmetric top molecules BaCH$_3$ and YbCH$_3$ have long-lived $K$-doublets accessible to experimental measurement with an even smaller splitting than in the linear polyatomic BaOH and YbOH molecules, typically $\leq 1$ MHz \cite{kozyryev2017precision}. The use of these $K$-doublets to measure the eEDM in BaCH$_3$ and YbCH$_3$ makes it possible to access large enhancement of the eEDM using moderate electric fields, reach a high experimental sensitivity, and avoid many systematic effects.

To calculate the $W_\mathrm{d}$ and $W_\mathrm{s}$ parameters, we employ high accuracy single reference and Fock-space coupled cluster methods and we explore the effect of the different computational factors on the calculated values. The employed methodology allows us to estimate the uncertainty in the calculated values of $W_\mathrm{d}$ and $W_\mathrm{s}$. The accurate and robust computational scheme established in this work may be extended to other polyatomic molecules sensitive to parity-violating effects. 


\section{Methodology}

The eEDM operator, $H^{\text{eEDM}}$, can be written in terms of a one-body operator  \cite{maartensson1987calculations} (here, and throughout the rest of this section, atomic units are used), 
\begin{equation}
    H^{\text{eEDM}}=2icd_\mathrm{e} \sum_i \gamma_i^5 \gamma_i^0\vv{p}_i^2,
\label{eq:H_de}    
\end{equation} 
where $\gamma^0, \gamma^1, \gamma^2$, and $\gamma^3$ represent the Dirac matrices,  $\gamma^5=i\gamma^0\gamma^1\gamma^2\gamma^3$, $\vv{p}_i$ is the momentum of the electron $i$, and $c$ is the speed of light. 

The S-PS interaction can be expressed as
\begin{equation}
   H^{\text{S-PS}}=i \frac{G_\mathrm{F}}{\sqrt{2}}Z_N k_\mathrm{s} \sum_i \gamma_i^0 \gamma_i^5\rho(\vv{r}_{iN}),
   \label{eq:H_ks-ps}
\end{equation}
where $G_\mathrm{F}$ is the Fermi constant ($2.2225\times10^{-14}$ a.u.), $Z_N$ is the atomic number of the nucleus $N$, and $\rho(\vv{r}_{iN})$ is the nuclear charge distribution. In equations (\ref{eq:H_de}) and (\ref{eq:H_ks-ps}), $d_\mathrm{e}$ and $k_\mathrm{s}$ parameterize the eEDM and S-PS interaction, respectively.

To calculate the electronic structure constants $W_\mathrm{d}$ and $W_\mathrm{s}$ using the coupled cluster approach, we use the finite field method  \cite{cohen1965electric,visscher1998molecular}, similar to our earlier works  \cite{haase2021systematic,denis2019enhancement,hao2018nuclear,norrgard2020nuclear,haase2020hyperfine,denis2022benchmarking}. 

The total Hamiltonian $H$ is expressed as a sum of a zeroth-order Hamiltonian $H^{(0)}$ and a perturbation $H_k$, regulated by the field strength parameter $\lambda_k$,
\begin{equation}
   H=H^{(0)} + \lambda_k H_k.
\end{equation}
In our case, $H^{(0)}$ is the unperturbed molecular Dirac-Coulomb Hamiltonian, 
\begin{equation}
    H^{(0)}=\sum_i[\beta_imc^2 + c \vv{\alpha}_i \cdot \vv{p}_i-V_{\text{nuc}}(\vv{r}_i)],
\end{equation}
where $\vv{\alpha}_i$ and $\beta_i$ are the Dirac matrices and $V_{\text{nuc}}$ is the Coulomb potential. Considering the perturbations $H^{\text{eEDM}}$ and $H^{\text{S-PS}}$,
\begin{equation}
    H_k=\frac{H^{\text{eEDM}}}{d_\mathrm{e}},\ \ \frac{H^{\text{S-PS}}}{k_\mathrm{s}},
\end{equation}
and applying the Hellman--Feynman theorem, the $W_\mathrm{d}$ and $W_\mathrm{s}$ coupling constants can be calculated as the first derivatives of the energy with respect to $\lambda_k$,
\begin{equation}
    W_\mathrm{d}=\frac{1}{\Omega}\bra{\Psi_{\Omega}^{(0)}}\frac{H^{\text{eEDM}}}{d_\mathrm{e}}\ket{\Psi_{\Omega}^{(0)}}= \frac{1}{\Omega}\left.\frac{dE_{\Omega}(\lambda_{d_\mathrm{e}})}{d \lambda_{d_\mathrm{e}}}\right|_{\lambda_{d_\mathrm{e}}=0}
\label{eq:derivative1}
\end{equation}
and
\begin{equation}
    W_\mathrm{s}=\frac{1}{\Omega}\bra{\Psi_{\Omega}^{(0)}}\frac{H^{\text{S-PS}}}{k_\mathrm{s}}\ket{\Psi_{\Omega}^{(0)}}=
    \frac{1}{\Omega} \left.\frac{dE_{\Omega}(\lambda_{k_\mathrm{s}})}{d \lambda_{k_\mathrm{s}}}\right|_{\lambda_{k_\mathrm{s}}=0},
\label{eq:derivative2}
\end{equation}
where $\Psi_{{\Omega}}^{(0)}$ is the unperturbed ground state electronic molecular wave function and $\Omega$ is the projection of the total electronic angular momentum  on the internuclear axis. 

The $W_\mathrm{d}$ and $W_\mathrm{s}$ constants are combined in the P,T-odd effective Hamiltonian, $H^{\text{P,T-odd}}$ \cite{kozlov1995parity}, 
\begin{equation}
    H^{\text{P,T-odd}} = (W_\mathrm{d} d_\mathrm{e} + W_\mathrm{s} k_\mathrm{s}) \vv{S}\cdot\vv{n}, 
\end{equation}
where $\vv{S}$ is the effective spin and $\vv{n}$ is a unit vector oriented along the internuclear axis. Therefore, a measurement of the P,T-violating energy difference on a single molecule provides us with the combined $d_\mathrm{e}$ and $k_\mathrm{s}$ and the disentanglement of the two effects requires experiments on different molecules  \cite{chupp2019electric}.


\section{Results}

In this work we assay a cost-accuracy balanced methodology that allows us to study polyatomic molecules promising for precision experiments, such as BaCH$_3$ and YbCH$_3$, at the high accuracy coupled cluster level. We study the effect of the treatment of relativity and electron correlation and the choice of the basis set on the optimised molecular geometries (section \ref{sec:geometry}) and on the calculated $W_\mathrm{d}$ and $W_\mathrm{s}$ parameters (section \ref{sec:enhancement}). Finally, we estimate the uncertainty of the predicted $W_\mathrm{d}$ and $W_\mathrm{s}$ based on an extensive computational study within the presented methodology (section \ref{sec:uncertainty}).

All the calculations were carried out using a modified version of the Dirac 2019 program \cite{DIRAC19,saue2020dirac}, except for the scalar relativistic (SR) calculations, where the CFOUR program \cite{cfour,matthews2020coupled} was employed. If not stated otherwise, the default settings of the corresponding codes were used. We applied both the single reference coupled cluster with single, double and perturbative triple excitations (CCSD(T))  \cite{visscher1996formulation} and the Fock-space coupled cluster (FSCC) with single and double excitations approach using sector (0,1) \cite{visscher2001formulation}. We used the uncontracted Dyall's relativistic basis sets  \cite{dyall2009relativistic,gomes2010relativistic,dyall2016relativistic} and the contracted atomic natural orbital correlation consistent ANO-RCC basis sets  \cite{roos2004relativistic,roos2004main,roos2008new,widmark1990density} of double-, triple-, quadruple-, and quintuple-zeta (for the ANO-RCC basis sets only) cardinality.

\subsection{Geometry optimisation}\label{sec:geometry}

The spectroscopic properties and the geometries of the polyatomic molecules that are considered to be good candidates for measurements of P,T-violating phenomena are usually not known \textit{a priori}, meaning that any computational study of these systems should begin with a geometry optimisation. 

Due to the large number of electrons, geometry optimisation of polyatomic molecules at the 4-component coupled cluster level requires impractically high computational resources. We thus look for a compromise that allows sufficiently accurate calculations at a realistic computational cost. In the Appendix \ref{sec:Appendix-opt} we describe the effects of the relevant computational parameters on the optimised geometry and use these investigations to select the optimal computational approach. After evaluating the effect of treatment of relativity, use of contracted vs. uncontracted basis set, and electron correlation, we conclude that a suitable methodology for reliable geometry optimisations of small polyatomic molecules is the combination of the SR-CCSD/CCSD(T) approach with the contracted ANO-RCC basis sets and correlating the valence and core-valence ($(n-1)$ and $(n-2)$) electrons. 

\begin{figure} 
    \centering
    \hspace{80pt}
    \includegraphics[scale=0.07]{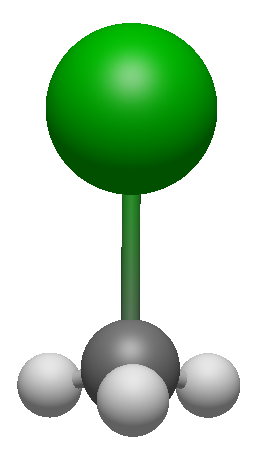}\\
    \vspace{-40pt}
    \includegraphics[scale=0.55]{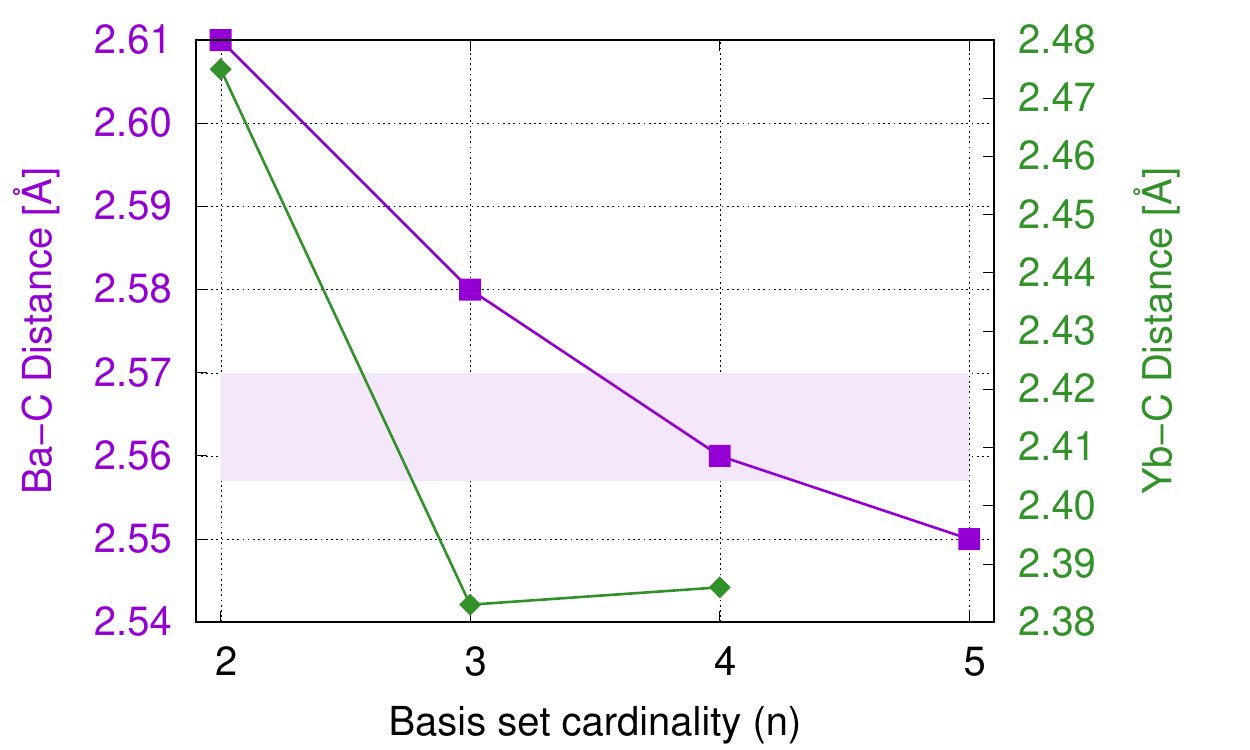}
    \caption{Convergence of the $M$--C bond length ($M$: Ba, Yb) with basis set cardinality (ANO-RCC-V$n$ZP basis sets). The calculations were carried out at the SR-CCSD(T) level of theory and correlating 37 and 51 electrons, respectively. The shaded area corresponds to the experimental bond range distance Ba--C in BaCH$_3$.}
    \label{fig:opt-basis_cardinality}
\end{figure}

Use of large basis sets is indispensable for taking full advantage of high accuracy correlation methods such as coupled cluster and for getting a good quality description of the system. Table \ref{tab:opt-basis_cardinality}  and Figure \ref{fig:opt-basis_cardinality} show the convergence of the obtained geometry of the two molecules with the increase of the cardinality of the ANO-RCC basis set. The optimised geometry of BaCH$_3$ (converged at the 5z basis set level) is close (0.3 -- 0.7\% difference) to the experimental geometry estimated from rotational constants in a gas-phase detection experiment (Ba--C bond distance 2.557 -- 2.570~\AA, BaCH angle 103$^\circ$ -- 108$^\circ$, both assuming C--H bond distance 1.1~\AA)  \cite{xin1998high}. This slight underestimation is most likely due to the combined effects of scalar relativity and basis set contraction, as can be seen from Table~\ref{tab:opt-hamiltonian_basis} in Appendix \ref{sec:Appendix-opt}. We evaluate the effect of this discrepancy on the $W_\mathrm{d}$ and $W_\mathrm{s}$ parameters and include it as a source of uncertainty in our reported values (see Section \ref{sec:uncertainty}).

In YbCH$_3$, we found a faster convergence than in BaCH$_3$, and thus used the geometry obtained with the QZ basis set. The optimised geometries used for the calculations of the $W_\mathrm{d}$ and $W_\mathrm{s}$ parameters in the following sections are presented in bold font in Table \ref{tab:opt-basis_cardinality}.

\begin{table} 
\centering
    \begin{tabular}{llll}
      \hline\hline
$n$ & Ba--C [\AA] & C--H [\AA] & BaCH [$^\circ$] \\
    \hline
2   &  2.61  & 1.11  & 113  \\ 
3   &  2.58  & 1.10  & 113  \\ 
4   &  2.56  & 1.09  & 113  \\ 
\bf{5} &  \bf{2.55}  & \bf{1.10}  & \bf{113} \\
Expt. \cite{xin1998high} & 2.557 -- 2.570 & 1.1 & 103 -- 108 \\
\hline
$n$ & Yb--C [\AA] & C--H [\AA] & YbCH [$^\circ$]\\
\hline
2 & 2.47 & 1.11 & 112 \\
3 & 2.38 & 1.09	& 111 \\
\bf{4} & \bf{2.39} & \bf{1.09}	& \bf{112} \\
\hline

\end{tabular}
\caption{Effect of basis set cardinality (ANO-RCC-V$n$ZP basis sets) on the optimised geometry of BaCH$_3$ and YbCH$_3$ at the SR-CCSD(T) level of theory and correlating 37 and 51 electrons, respectively. The final optimised geometries are highlighted in bold. }
\label{tab:opt-basis_cardinality}
\end{table}

\subsection{Enhancement factors: Computational parameters}\label{sec:enhancement}

Both the $W_\mathrm{d}$ and $W_\mathrm{s}$ factors  are purely relativistic properties and, according to Schiff's theorem, in the non-relativistic regime the atom or the molecule would not have an EDM even if the electron did  \cite{schiff1963measurability}. However, Schiff's theorem is not valid in the relativistic regime, and atoms and molecules may express a non-zero EDM  \cite{sandars1965electric}. Therefore, calculations of the $W_\mathrm{d}$ and $W_\mathrm{s}$ factors should be carried out in a relativistic framework. 

In this work, we use the relativistic Dirac--Coulomb 4-component Hamiltonian combined with the single-reference coupled cluster approximation  \cite{visscher1996formulation} for the calculation of the $W_\mathrm{d}$ and $W_\mathrm{s}$ factors in BaCH$_3$, and the multireference Fock-space coupled cluster approach \cite{visscher2001formulation}, for the corresponding calculations in YbCH$_3$. 

Earlier calculations of the electric-field gradients in YbF  \cite{pavsteka2016relativistic} and of $W_\mathrm{d}$ in YbF \cite{haase2021systematic} and YbOH  \cite{denis2019enhancement} have shown that Yb-containing molecules may present a multireference character in their ground states. In this work, we found, through the T1 diagnostic  \cite{lee1989diagnostic}, that the ground state of YbCH$_3$ does, indeed, could indeed benefit from Fock space coupled cluster method. In the FSCC(0,1) approach, the starting point is the closed shell ground state of YbCH$_3^+$ molecule to which one electron is added within the correlation procedure. In this work, we added the electron to the lowest energy $\sigma$ orbital, thus using a minimal model space. The consideration of higher energy orbitals through a larger model space was not shown to have a significant effect on the value of $W_\mathrm{d}$ in YbF and YbOH \cite{denis2019enhancement,haase2021systematic}.

We apply equations \eqref{eq:derivative1} and \eqref{eq:derivative2} for the calculations of both the $W_\mathrm{d}$ and $W_\mathrm{s}$ factors, using $\lambda_{d_\mathrm{e}}= 10^{-8}$~a.u. and $\lambda_{k_\mathrm{s}}=  10^{-7}$~a.u. for BaCH$_3$ and $\lambda_{d_\mathrm{e}}=\lambda_{k_\mathrm{s}}= 10^{-6}$~a.u. for YbCH$_3$. These field strengths were selected guarantying numerical stability as it is shown in Appendix \ref{sec:Appendix-ffpt}.
In all cases, the convergence criteria of the coupled cluster energy as well as the Hartree--Fock energy was fixed in the $1\times10^{-11} - 5\times10^{-11}$~a.u range. 

In the following, we present the recommended $W_\mathrm{d}$ and $W_\mathrm{s}$ values and their uncertainties and discuss the scheme we use to determine the latter, focusing separately on the various parameters that determine the quality of the calculations. Initially, we focus our study on $W_\mathrm{d}$, and subsequently, we include $W_\mathrm{s}$ in our discussion. 

\subsubsection{Electron correlation}

In this work, we investigate the effect of various computational parameters within the  relativistic coupled cluster approach on the calculated $W_\mathrm{d}$ and $W_\mathrm{s}$ factors.

\begin{figure} 
    \centering
    \includegraphics[scale=0.55]{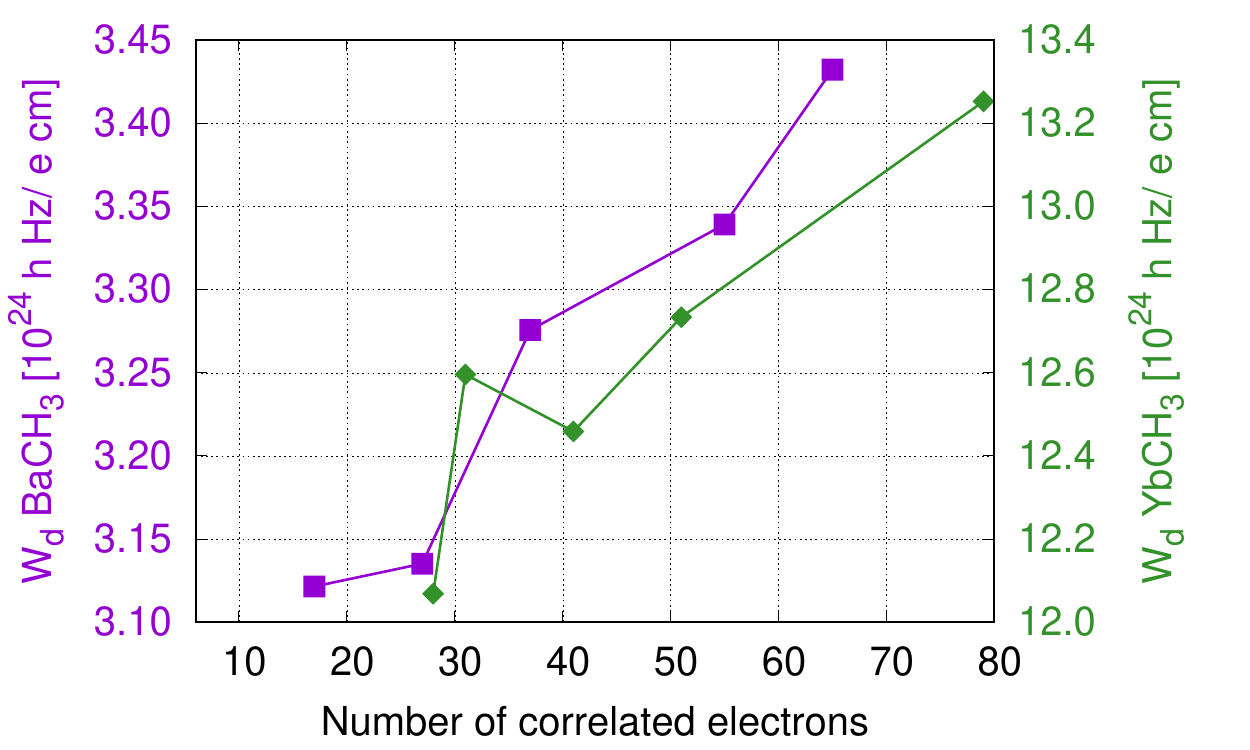}
    \caption{Effect of the number of correlated electrons, $N$, on the calculated $W_\mathrm{d}$ constants. Values presented in Table \ref{tab:Wd-ecorr}.}
    \label{fig:Wd-ecorr}
\end{figure}

\paragraph{Correlation space}
Table \ref{tab:Wd-ecorr} and Figure \ref{fig:Wd-ecorr} present the CCSD results obtained correlating a different number of electrons; in all these calculations, the virtual cut-off was adjusted symmetrically to the number of electrons correlated (that is the positive energy cut-off was taken to be of the same absolute size as the negative energy cut-off). 
We observe that correlating only the outer-core-valence electrons (17 electrons in BaCH$_3$; Ba: 5s; 5p; 6s, C: 2s; 2p, H: 1s, and 29 electrons in YbCH$_3$; Yb: 5p; 4f; 6s, C: 2s; 2p, H: 1s) causes relative errors of $\sim$9\% compared to the value obtained when all the electrons are included in the correlation treatment. 
Previously, we found that correlating only outer-core-valence electrons in  the isoelectronic BaF also led to an error of $\sim$10\%  compared to the all-electon calculation  \cite{haase2021systematic}.

Therefore, for the recommended values, all the electrons were correlated.
Note that the strong dependence on the description of the core region is unusual for most other atomic and molecular properties, such as molecular geometries or spectra. 

\begin{table} 
\centering
\begin{tabular}{clll}
\hline \hline
$N$ & \multicolumn{2}{c}{Frozen orb.} &$W_\mathrm{d}$ [$10^{24}\frac{h\text{Hz}}{e\text{cm}}$] \\
& Ba & C & \\
\hline
    17 & [Kr]4d & [He]&3.12 (--9.1\%) \\  
    27 & [Kr]   & [He]&3.14 (--8.7\%) \\  
    37 & [Ar]3d & --  &3.28 (--4.6\%) \\  
    55 & [Ne]   & --  &3.34 (--2.7\%) \\  
    65 & --     & --  &3.43 \\
    \hline
 
    $N$ & \multicolumn{2}{c}{Frozen orb.} &\text{$W_\mathrm{d}$ } [$10^{24}\frac{h\text{Hz}}{e\text{cm}}$]\\
     & Yb & C &\\
    29 & [Kr]4d5s & [He]  & 12.07 (--8.9\%)\\
    31 & [Kr]4d & [He]& 12.60 (--5.0\%)\\
    41 & [Kr]   & [He]& 12.46 (--6.0\%)\\
    51 & [Ar]3d & -- & 12.73 (--3.9\%)\\
    79 & --     & --  & 13.25 \\
\hline
\end{tabular}
\caption{Effect of the number of correlated electrons, $N$, on the calculated $W_\mathrm{d}$ constants. Error relative to the all-electrons-correlated result is presented in parenthesis. Relativistic CCSD and FSCC approaches for BaCH$_3$ and YbCH$3$, respectively, were used, combined with the dyall.v2z basis sets.}
\label{tab:Wd-ecorr}
\end{table}

\begin{figure} 
    \centering
    \includegraphics[scale=0.55]{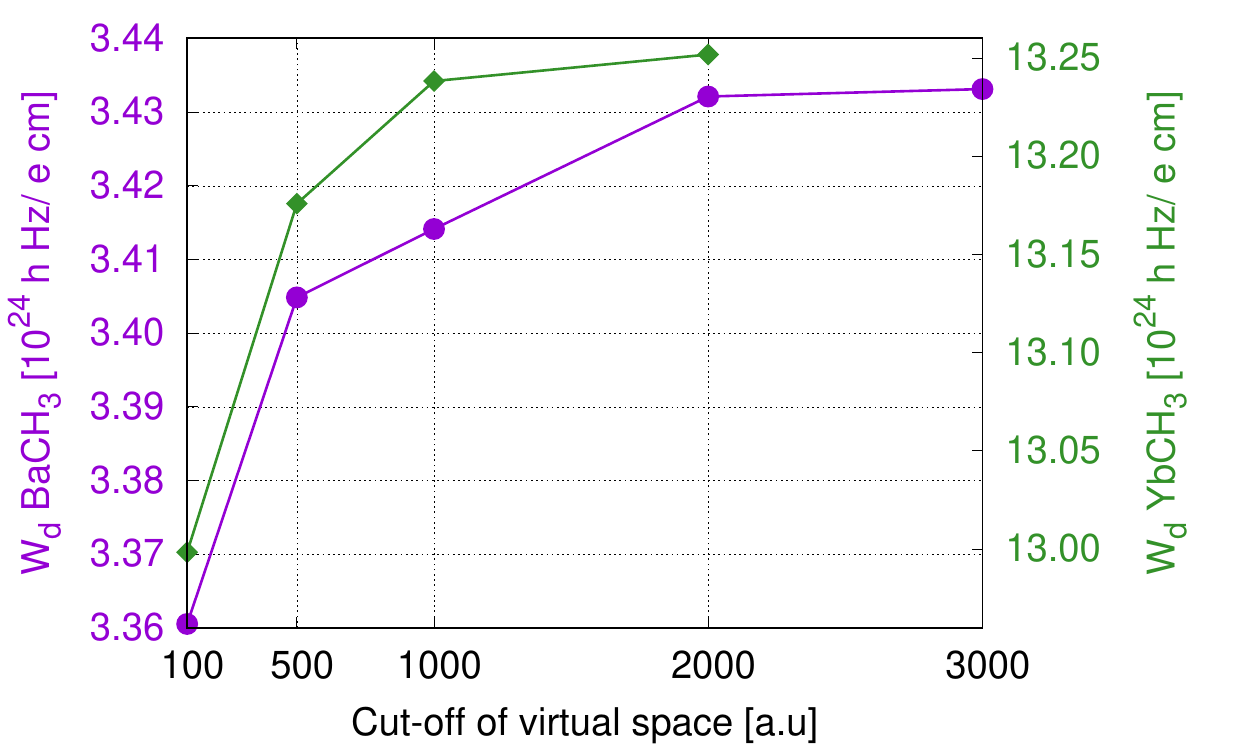}
    \caption{Effect of the virtual space cut-off on the calculated $W_\mathrm{d}$ constants. Values presented in Table \ref{tab:Wd-virtual_cutoff}.}
    \label{fig:Wd-virtual_cutoff}
\end{figure}

Correlation of all the electrons in the coupled cluster approach requires simultaneous inclusion of a proportionally large number of virtual orbitals. Notice that the symmetric virtual cut-off is $\sim1400$ a.u. in BaCH$_3$ and $\sim2300$ a.u. in YbCH$_3$. Figure \ref{fig:Wd-virtual_cutoff} and Table \ref{tab:Wd-virtual_cutoff} present the effect of the virtual cut-off on the calculated $W_\mathrm{d}$, where all the electrons are correlated.  According to the results obtained using  virtual cut-offs of 1000 and 2000 a.u., increasing the cut-off over 1000 a.u. changes the reported $W_\mathrm{d}$ by less than 0.5\% for BaCH$_3$ and by less than 0.1\% for YbCH$_3$. In fact, $W_\mathrm{d}$ in BaCH$_3$ changes by only 0.03\% when the virtual cut-off is further increased to 3000 a.u. Therefore, we include all the virtual orbitals until 2000 a.u. in our recommended values, and we use the difference between the results obtained with a virtual cut-off of 2000 and of 1000 a.u. to estimate the uncertainty from neglecting virtual orbitals above  2000 a.u. in the description of the electron correlation. Furthermore, we conclude that when computational resources are a bottleneck, lower cut-off of 500 a.u. can be used, without notable deterioration of the quality of the results. 

\begin{table} 
\centering
\begin{tabular}{lll}
\hline \hline
Cut-off& \multicolumn{2}{c}{$W_\mathrm{d}$  [$10^{24}\frac{h\text{Hz}}{e\text{cm}}$]}\\
\cline{2-3}
[a.u.] &  BaCH$_3$   &  YbCH$_3$ \\
    \hline
     100 & 3.36 (--2.1\%) & 13.00 (--1.9\%)\\
     500 & 3.41 (--0.8\%) & 13.18 (--0.6\%)\\
    1000 & 3.42 (--0.5\%) & 13.24 (--0.1\%)\\
    2000 & 3.43 (--0.03\%) & 13.25 \\
    3000 & 3.43 & -- \\
    \hline
\end{tabular}
\caption{Effect of the virtual space cut-off on the calculated $W_\mathrm{d}$ constants. Error relative to the highest virtual cut-off value is presented in parenthesis. Relativistic CCSD and FSCC approaches for BaCH$_3$ and YbCH$3$, respectively, were used, combined with the dyall.v2z basis sets. All electrons were correlated.}
\label{tab:Wd-virtual_cutoff}
\end{table}

\paragraph{Excitation rank}
The results discussed so far were obtained at the coupled cluster (either single reference or Fock-space) level of theory with single and double excitations. 

We evaluated the effect of triple excitations comparing the calculated $W_\mathrm{d}$ of BaCH$_3$ at the CCSD and CCSD(T) levels using the 2z, 3z, and 4z quality basis sets (Table~\ref{tab:Wd-triples}).
In all the cases, the inclusion of triple excitations slightly reduces the calculated $W_\mathrm{d}$.
The contribution of the triple excitations increases with the cardinality of the basis set, as expected, but even at the 4z basis set cardinality reaches only 1.3\%. The magnitude of this contribution is close to what was observed in previous calculations on BaOH (1.4\%) and BaF (1.5\%) and suggests that the triple excitations can be neglected when the computational resources are an issue. 
We report the final $W_\mathrm{d}$ and $W_\mathrm{s}$ at the CCSD(T) level and use the triples contribution to estimate the uncertainty due to neglecting higher-order excitations.

The use of the FSCC approach allows us to obtain accurate values of $W_\mathrm{d}$ and $W_\mathrm{s}$ in a open shell system like YbCH$_3$. However, as perturbative triple excitations are not yet implemented in the FSCC module of the DIRAC program, we can not use this method to perform a reliable calculation of the contribution of the triple and higher-order excitations. Therefore, we do use the difference between the single reference CCSD and CCSD(T) results in BaCH$_3$ to estimate the relative uncertainty due to the neglect of the triple excitations in YbCH$_3$ molecule. 

\begin{table} 
\centering
\begin{tabular}{ccc}
\hline \hline
\multirow{2}{*}{Basis set} & \multicolumn{2}{c}{ $W_\mathrm{d}$ BaCH$_3$} [$10^{24}\frac{h\text{Hz}}{e\text{cm}}$]\\
\cline{2-3}
 &  \text{CCSD} &  \text{CCSD(T)} \\
    \hline
    dyall.v2z & 3.12 (+0.6\%) & 3.10 \\
    dyall.v3z & 3.01 (+1.0\%) & 2.98 \\
    dyall.v4z & 2.95 (+1.3\%) & 2.91\\
    \hline
\end{tabular}
\caption{Effect of inclusion of triple excitations on the calculated $W_\mathrm{d}$ constants. The relative error of CCSD comparing to the perturvative CCSD(T) is presented in parenthesis. 17 electrons were correlated and the virtual cut-off was set to 30 a.u.}
\label{tab:Wd-triples}
\end{table}

\subsubsection{Basis sets}

Next to the electron correlation, the quality of the basis set plays a crucial role in accurate theoretical description of molecular properties. In the following subsections we analyse the effect of the cardinality and the special features of the basis sets on the calculated $W_\mathrm{d}$ and $W_\mathrm{s}$ factors.

\paragraph{Complete basis set limit}

Table \ref{tab:Wd-cardinality_basis} presents the dependence of the calculated $W_\mathrm{d}$ constants on the cardinality of the basis set. Overall, $W_\mathrm{d}$ changes monotonically with increase of basis set cardinality. In BaCH$_3$ it decreases by 6.5\%  when going from 2z to 4z cardinality basis set, while In YbCH$_3$ it increases by 4.1\%. We extrapolate our results to the complete basis set (CBS) limit, using the usual three-point Dunning--Feller $e^{-\alpha n}$ scheme ($n=2,3,4$) \cite{dunning1989gaussian, feller1992application} for extrapolating the DHF energies and the two-point Helgaker et al. $n^{-3}$ scheme ($n=3,4$) \cite{helgaker1997basis} for extrapolating the correlation energies. We also tested the Martin $(n+\frac{1}{2})^{-4}$ scheme  \cite{martin1996ab} and the recent scheme of Lesiuk and Jeziorski (based on the application of the Riemann zeta function)  \cite{lesiuk2019complete} for extrapolating the correlation energies. The latter three extrapolation schemes (labeled H, M and L in Table~\ref{tab:Wd-cardinality_basis}, respectively) give evenly spread and very similar CBS limits. We use the central Helgaker CBS limit for our final results and the respective 95\% confidence interval ($1.96\sigma$) based on the spread of the three schemes as our extrapolation uncertainty estimate. The same methodology was used in our previous studies \cite{guo2021ionization}.

\begin{table} 
\centering
\begin{tabular}{lll}
\hline \hline
\multirow{2}{*}{Basis set}& \multicolumn{2}{c}{$W_\mathrm{d}$  [$10^{24}\frac{h\text{Hz}}{e\text{cm}}$]}\\
\cline{2-3}
 &  BaCH$_3$   &  YbCH$_3$ \\
    \hline
    dyall.v2z & 3.10 $(+8.3\%)$ & 12.07 $(-4.4\%)$  \\
    dyall.v3z & 2.98 $(+4.1\%)$ & 12.52 $(-0.8\%)$  \\
    dyall.v4z & 2.91 $(+1.7\%)$ & 12.58 $(-0.4\%)$  \\
    \hline
    CBS(M) & 2.87 $(+0.3\%)$ & 12.62 $(-0.1\%)$ \\
    CBS(H) & 2.86 & 12.63 \\
    CBS(L) & 2.85 $(-0.4\%)$ & 12.64 $(+0.1\%)$ \\
\hline
95\% c.i. & 0.02 & 0.02\\
    \hline
\end{tabular}
\caption{Effect of the basis set cardinality (v$n$z, $n=2,3,4$) on the $W_\mathrm{d}$ constants calculated using CCSD(T) with 17  correlated electrons and a virtual space cut-off of 30 a.u. for BaCH$_3$ and using FSCC with 29 electrons correlated and virtual space cut-off of 10 a.u. for YbCH$_3$.
In the lower part of the table, results obtained using the different CBS extrapolation schemes and the respective uncertainty estimation are shown. Relative errors with respect to the CBS(H) limit are shown in parentheses.}
\label{tab:Wd-cardinality_basis}
\end{table}

\paragraph{Core correlating and diffuse functions}

The results shown in Table~\ref{tab:Wd-aumented_basis} showcase the effects of including outer- and inner-core correlating (cv3z, aev3z) and diffuse functions (aug-v3z) on the calculated $W_\mathrm{d}$ factors. The difference between the cv3z/ae3z basis set and the v3z basis set lies in the addition of tight functions with high angular momentum, namely, 2/5~f and 1/2~g~functions for Ba atom and 1/1~d~function for C. On the other hand, the difference between the aug-v3z and the v3z lies in the even-tempered addition of diffuse functions for each angular momentum in all the atoms. For BaCH$_3$ the two augmentation schemes cause changes that are similar in magnitude but opposite in sign, leading to a negligible net effect. The $W_\mathrm{d}$ in YbCH$_3$ shows a negligible dependence on both types of augmentation. In this case, the difference between the dyall.v3z and the dyall.cv3z basis sets is small since the two sets are identical for Yb and the only differences are in the number of the core-correlating functions on carbon. In the dyall.ae3z basis set for Yb, only one extra g function is added to the dyall.cv3z basis. Due to the small (and opposite) effects, we proceed with the non-augmented dyall.v3z basis sets for both molecules and we included the effect of the augmentation and including correlating functions in our uncertainty estimation. 

\begin{table} 
\centering
\begin{tabular}{lll}
\hline \hline
& \multicolumn{2}{c}{$W_\mathrm{d}$ [$10^{24}\frac{h\text{Hz}}{e\text{cm}}$]}\\
 \cline{2-3}
\text{Basis set} &  BaCH$_3$  &   YbCH$_3$ \\
    \hline
    dyall.v3z & 3.20  & 12.87  \\
    dyall.cv3z & 3.26 $(+1.71\%)$& 12.88 $(+0.05\%)$\\ 
    dyall.ae3z & 3.26 $(+1.68\%)$ & 12.86 $(-0.08\%)$\\
    aug-dyall.v3z & 3.15 $(-1.74\%)$ & 12.87 $(-0.02\%)$\\
    \hline
\end{tabular}
\caption{Effect of the basis set on the calculated $W_\mathrm{d}$ constants. The deviations relative to the dyall.v3z basis set are shown in parenthesis. CCSD(T) level of theory with 37 electrons correlated and virtual space cut-off of 60 a.u. for BaCH$_3$ and FSCC with 31 electrons correlated and virtual space cut-off of 20 a.u. for YbCH$_3$.}
\label{tab:Wd-aumented_basis}
\end{table}


\subsection{Recommended values and uncertainty estimation}
\label{sec:uncertainty}

\begin{table*} 
\centering
\begin{tabular}{lllll}
\hline\hline
 \multicolumn{2}{l}{\multirow{2}{*}{Source of uncertainty}} & \multicolumn{2}{c}{$\delta_i$ $W_\mathrm{d}$ [$10^{24}\frac{h\text{Hz}}{e\text{cm}}$] ($\%$)}  & \multirow{2}{*}{Scheme}   \\
 \cline{3-4}
 &&BaCH$_3$&YbCH$_3$ &\\
\hline
\multirow{2}{*}{Correlation}  & Virtual cut-off & 0.017 (0.5\%) &0.014 (0.1\%)& 2000 -- 1000 a.u.   \\
& Triples & 0.029 (0.9\%) & 0.252 (1.8\%) & CCSD(T) -- CCSD  \\
 \hline
\multirow{3}{*}{Basis set} & CBS extrapolation & 0.021 (0.7\%) & 0.020 (0.15\%)& 1.96$\sigma$ \\
 & Diffuse functions & 0.055 (1.7\%) &0.003 (0.02\%)& aug-v3z -- v3z   \\
 & Core-correl. functions  & 0.055 (1.7\%) &0.010 (0.07\%) & ae3z -- v3z  \\
 \hline
 Geometry & Optimization & 0.037 (1.2\%) & 0.160 (1.2\%) & opt. -- expt.\\
 \hline
 Relativity & Gaunt & 0.057 (1.8\%) & 0.180 (1.3\%) & Gaunt-DC -- DC\\
 \hline
 \multicolumn{2}{l}{ Total uncertainty} & 0.111 (3.44\%) & 0.349 (2.53\%) & $\sqrt{\sum_i \delta_i^2}$\\
\multicolumn{2}{l}{ Final recommended value}  &3.224 $\pm$ 0.111  & 13.799 $\pm$ 0.349  & \\
\hline
\end{tabular}
\caption{Summary of the most significant sources of uncertainty in $W_\mathrm{d}$ in BaCH$_3$ and YbCH$_3$ [$10^{24}\frac{h\text{Hz}}{e\text{cm}}$]. Values in parentheses represent the relative uncertainties with respect to the final results.}
\label{tab:uncertainty}
\end{table*}  

The extensive computational study carried out in the previous section allows us to determine the most suitable method for obtaining the recommended values of the $W_\mathrm{d}$ and the $W_\mathrm{s}$ constants of the two molecules. 

For BaCH$_3$, we provide the final value of $W_\mathrm{d}$ calculated at the CCSD(T) level of theory using the dyall.v3z basis set, correlating all the electrons and including virtual orbitals up to 2000~a.u. giving the base value of $3.33\times 10^{24}\frac{h\text{Hz}}{e\text{cm}}$. To this, we add the CBS extrapolation correction of $-0.11\times 10^{24}\frac{h\text{Hz}}{e\text{cm}}$ evaluated correlating 17~electrons with a virtual space cut-off of 30~a.u. We estimated the uncertainty due to the finite basis set (cardinality, core correlating and diffuse basis functions) and due to the neglect of higher-order excitations based on the study in the previous sections. To determine the size of each source of error, we use the difference in $W_\mathrm{d}$ obtained with the final method and $W_\mathrm{d}$ obtained with a lower approximation (for a given computational parameter) as is schematically shown in Table~\ref{tab:uncertainty}. In addition, to estimate the uncertainty stemming from the molecular geometry optimisation, we calculated the $W_\mathrm{d}$ factor using the experimental geometry range (Ba--C bond distance = 2.557 -- 2.570~\AA \cite{xin1998high}) and found that the maximum variation with respect to the value obtained using the optimised geometry is 1.08\%. Finally, the Dirac--Coulomb Hamiltonian used in this work assumes an instantaneous (non-relativistic) electron-electron interaction. The Breit interaction is the first-order correction, and as it contains two-electron operators \cite{lindroth1989order} it is not possible to separate this effect from the electron correlation effects. While aware of this limitation, we evaluate effect of the dominant Gaunt term of the Breit interaction \cite{gaunt1929iv} 
at the  Dirac--Hartree--Fock (DHF) level of theory resulting in an decrease in $W_\mathrm{d}$ of 1.8\%. We thus set an estimate of the uncertainty on our recommended values due to the missing Breit interaction and higher order effects to 1.8\%. All the individual contributions to the uncertainty are summarised in Table~\ref{tab:uncertainty} and we estimate the total uncertainty (assuming the effects are independent) at 3.44\%. The recommended value for BaCH$_3$ is thus $W_\mathrm{d}=3.224\pm 0.111\times 10^{24}\frac{h\text{Hz}}{e\text{cm}}$. 

For YbCH$_3$, we provide the final value of $W_\mathrm{d}$ calculated at the FSCC level of theory, correlating all the electrons and including virtual orbitals up to 2000 a.u, as we did in BaCH$_3$. In YbCH$_3$ however, it was computationally unfeasible to correlate all the electrons when using the dyall.v3z basis. Nevertheless, since the effect of the correlation space is bigger than the effect of the size of the basis set, we employed the dyall.v2z basis set to obtain the base value of $13.25\times 10^{24}\frac{h\text{Hz}}{e\text{cm}}$ and to this we added the CBS correction of $+0.56\times 10^{24}\frac{h\text{Hz}}{e\text{cm}}$, evaluated correlating 29~electrons with a virtual space cut-off of 10~a.u. We used an analogous scheme to that employed for BaCH$_3$ to estimate the uncertainty of this result. However, since the experimental geometry is yet not known for YbCH$_3$, we use the same relative uncertainty due to the geometry optimisation as we derived for BaCH$_3$. This assumption is justified by the fact that we used the same methodology to optimise the geometry of the two molecules. Furthermore, to estimate the effect of the excitation rank, we include in our uncertainty estimation twice the relative uncertainty obtained for BaCH$_3$ to account for neglecting both the triples and the higher-order excitations. The recommended value for YbCH$_3$ is thus $W_\mathrm{d}=13.799\pm 0.349\times 10^{24}\frac{h\text{Hz}}{e\text{cm}}$.  

\begin{figure} 
    \centering
    \includegraphics[scale=0.55]{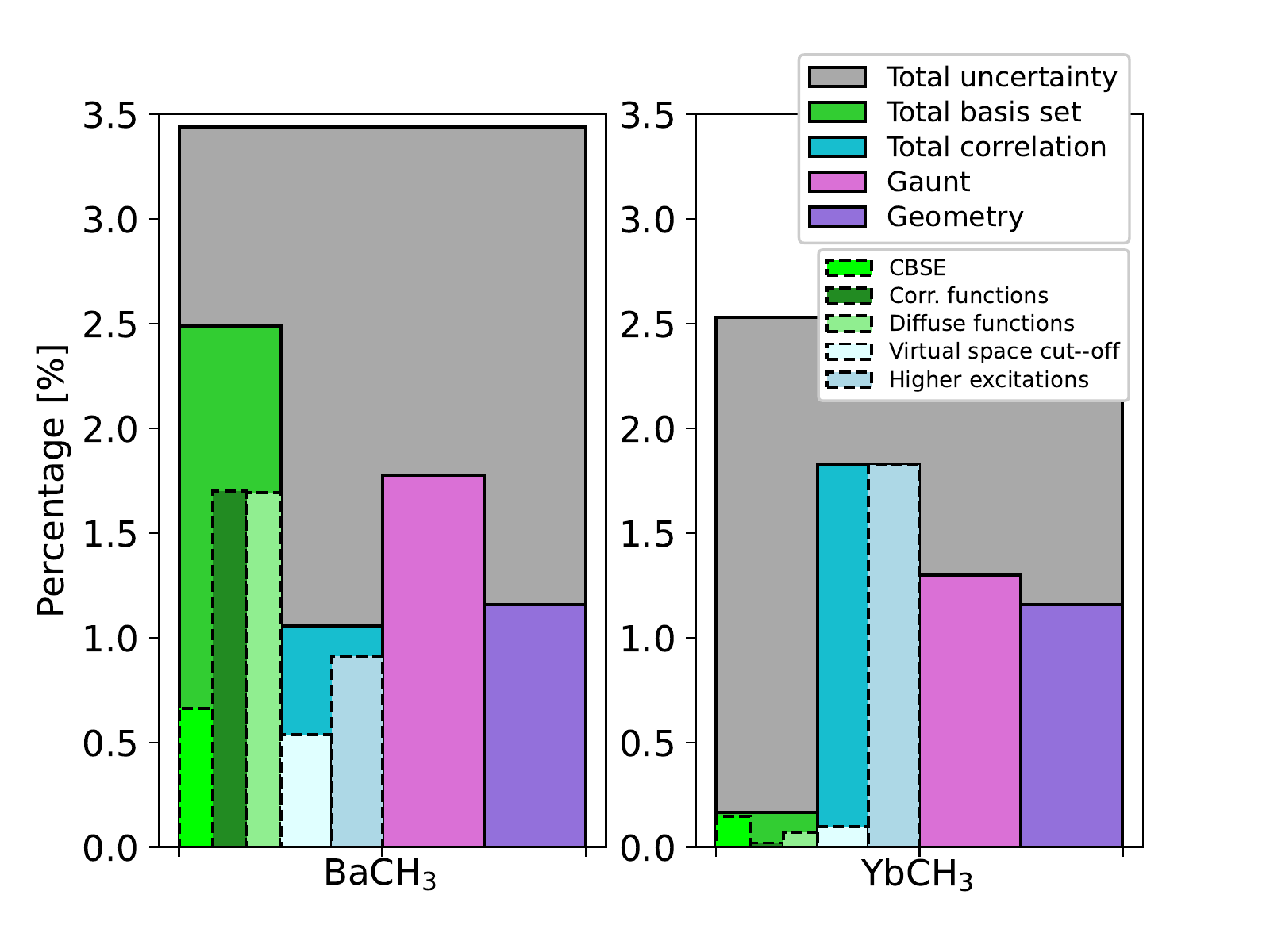}
    \caption{Total and individual contributions to the uncertainty in percent relative to the recommended values (DC-CCSD(T), CBSE-corrected, and DC-FSCC, CBSE-corrected for BaCH$_3$ and YbCH$_3$, respectivly).}
    \label{fig:uncertainty}
\end{figure}

Figure~\ref{fig:uncertainty} presents the relative uncertainties for the different sources discussed above and calculated as described in Table~\ref{tab:uncertainty}. The highest contributions to the total uncertainty in BaCH$_3$ is due to the basis set incompleteness and the current limitation to describing the Breit interaction correctly. Notice that the addition of diffuse and extra correlating functions have an opposite effect, meaning that we over-estimate somewhat this source of uncertainty. In the case of YbCH$_3$, the employed basis set is highly converged. Consequently, the total uncertainty in $W_\mathrm{d}$ is smaller for YbCH$_3$ than for BaCH$_3$ with the leading source of uncertainty coming from the electron correlation description.   

We have also performed calculations of the $W_\mathrm{s}$ parameters using the computational scheme employed for the recommended $W_\mathrm{d}$ values. The different computational parameters included in the uncertainty estimation in this work have effects of the same magnitude on the calculated $W_\mathrm{d}$ and $W_\mathrm{s}$ in BaF \cite{haase2021systematic}, with the Gaunt term being the only exception. In BaF, the Gaunt term, and therefore, the total uncertainty, is smaller in $W_\mathrm{s}$ than in $W_d$. Consequently, in this work we assume the relative uncertainty of the $W_\mathrm{s}$ values is on the same order of magnitude as the uncertainty found for the corresponding $W_\mathrm{d}$ factors. All the recommended values are summarised in Table~\ref{tab:comparision} and compared to the earlier predictions for similar molecules. 

\begin{table} 
\footnotesize
    \centering
    \begin{tabular}{llll}
        \hline \hline
        Molecule & Method & $W_\mathrm{d}$ [$10^{24}\frac{h\text{Hz}}{e\text{cm}}$] & $W_\mathrm{s}$ [$h$kHz] \\
        \hline
        BaCH$_3$ & DC-CCSD(T) & $\bf{3.22\pm0.11}$ &$\bf{8.42\pm 0.29}$\\
        BaOCH$_3$ & X2C-CCSD(T) & 3.05  \cite{zhang2021calculations} & \\
        BaOH     & DC-CCSD(T) & $3.10\pm0.15$ \cite{denis2019enhancement}&\\
         & ZORA-cGHF & $3.32\pm0.33$  \cite{gaul2020ab}& $8.79\pm0.88$ \cite{gaul2020ab} \\
         & ZORA-cGKS & $2.98\pm0.30$  \cite{gaul2020ab}& $7.91\pm0.79$ \cite{gaul2020ab} \\ 
        BaF     &  DC-CCSD(T) & $3.13\pm0.12$ \cite{haase2021systematic}&$8.29\pm0.12$ \cite{haase2021systematic}\\
        \hline
        YbCH$_3$ & DC-FSCC(0,1) &$\bf{13.80\pm0.35}$ & $\bf{45.35\pm1.15}$\\
        YbOCH$_3$ & X2C-CCSD(T) & 11.60  \cite{zhang2021calculations} & \\
        YbOH     & DC-FSCC(0,1) &$11.30\pm0.5$ \cite{denis2019enhancement} &\\
         & ZORA-cHFS & $11.40\pm 1.14$  \cite{gaul2020ab}& $41.2\pm 4.12$  \cite{gaul2020ab}\\
         & ZORA-cGKS & $8.54\pm 0.85$  \cite{gaul2020ab}& $30.8\pm 3.08$  \cite{gaul2020ab}\\
         & DC-CCSD & $11.47\pm 0.68$  \cite{prasannaa2019enhanced}& \\
        YbF & DC-FSCC    & 11.39 \cite{haase2021systematic}  &\\
         \hline
    \end{tabular}
    \caption{Recommended $W_\mathrm{d}$ and $W_\mathrm{s}$ values of  BaCH$_3$ and YbCH$_3$ and comparison to homologous molecules. A factor of 0.4835 was used to convert E$_{\text{eff}}$ in GV/cm units to $W_\mathrm{d}$ in $10^{24}\frac{h\text{Hz}}{e\text{cm}}$ units when necessary.}
    \label{tab:comparision}
\end{table}

\subsection{Comparison and molecular bond analysis} \label{comparison-bond}

The final calculated enhancement factors in BaCH$_3$ and YbCH$_3$ are of the same order of magnitude as in their corresponding isoelectronic linear molecules BaOH, BaF, YbOH, and YbF, and the non-linear polyatomic molecules BaOCH$_3$ and YbOCH$_3$, respectively. However, the enhancement factors in BaCH$_3$ and especially in YbCH$_3$ are larger than in the other molecules.

To investigate the origin of this difference, we calculated the $W_\mathrm{d}$ and $W_\mathrm{s}$ parameters for all the systems using the same approach and used the Quantum Theory of Atoms In Molecules (QTAIM) \cite{bader1981quantum} and the Natural Bond Orbital (NBO) \cite{glendening2013nbo} analyses to study the relevant bonds. Table \ref{tab:comparison-bond} presents $W_\mathrm{d}$ and $W_\mathrm{s}$ obtained at the CCSD(T) and FSCC(0,1) level for the Ba- and Yb-containing molecules, respectively. To conserve computational effort, the dyall.v2z basis set was used and a cutoff was set at --2 to +10~a.u. for Ba- and --1 to +10~a.u. for Yb-containing systems. For the QTAIM and NBO analysis, we employ the SR-ZORA Hamiltonian with the PBE functional~\cite{PhysRevLett.77.3865} and the QZ4P basis set in the Amsterdam density functional (ADF) package \cite{te2001chemistry}.

In the QTAIM analysis, intermolecular interactions can be characterized according to the topology of the electron density at the bond critical points (BCPs) denoted by $\vv{r}_c$. Specifically, the Espinoza and coworkers criteria \cite{espinosa2002weak} classify the bonding interactions according to the ratio of the potential $V(\vv{r}_c)$ and kinetic $G(\vv{r}_c)$ energy density at the BCPs, $|V(\vv{r}_c)|/G(\vv{r}_c)$. A ratios $<1$ correspond to long-range, ionic or hydrogen bonds, values between 1 -- 2  to intermediate bonds (with both ionic and covalent character) and ratios $>2$ to covalent bonds.
The ratio $|V(\vv{r}_c)|/G(\vv{r}_c)$ in Table \ref{tab:comparison-bond} suggests that the M--X  (M = Ba, Yb; X = C, O, F) bonds can not be characterized as purely ionic bonds, and especially, the M--C bonds have a significantly increased covalent character as compared to the M--O and M--F bonds. 
The natural charges based on natural atomic orbitals (NAOs) of the heavy atoms $q_\mathrm{M}$ support this observation. The M--C bonds are less polar compared to the M--O and M--F bonds. It means that in BaCH$_3$ and YbCH$_3$ molecules, the unpaired electron that experiences the P, T--odd interaction is more localized on the heavy atom, leading to a higher enhancement factor than for the other Ba-, Yb-containing molecules. 
We tested this conjencture by including YbH in our analysis, which follows the expected trend -- the less electronegative hydrogen leads to more covalent bonding and an increase in the $W_\mathrm{d}$ and $W_\mathrm{s}$ values.
A similar observation was made in previous works~\cite{sunaga2017analysis,meyer2006candidate}.

\begin{table}
\centering
\begin{tabular}{lcccc}
\hline\hline
\multirow{2}{*}{System} & $W_\mathrm{d}$ & $W_\mathrm{s}$ & \multirow{2}{*}{$\frac{|V(\vv{r}_c)|}{G(\vv{r}_c)}$} & \multirow{2}{*}{$q_\mathrm{M}$} \\
 & [$10^{24}\frac{h\text{Hz}}{e\text{cm}}$] & [$h$kHz] &  &   \\
 \hline
BaCH$_3$ & \phantom{0}3.10 & \phantom{0}7.60 & 1.64 & 0.78  \\
BaOCH$_3$ & \phantom{0}2.79 & \phantom{0}6.88 & 1.20 & 0.88  \\
BaOH & \phantom{0}2.82 & \phantom{0}6.96 & 1.22 & 0.90   \\
BaF & \phantom{0}2.80 & \phantom{0}6.89 & 1.28 & 0.89  \\
 \hline
YbCH$_3$ & 12.07 & 41.20 & 1.48 & 0.68 \\
YbOCH$_3$ & \phantom{0}9.89 & 33.98 & 1.12 & 0.82 \\
YbOH & 10.03 & 34.51 & 1.13 & 0.84 \\
YbF & 10.30 & 35.43 & 1.15 & 0.83  \\
YbH  & 13.15 & 45.05 & 1.55 & 0.64 \\
 \hline
 \end{tabular}
 \caption{Correlation between the $W_\mathrm{d}$ and $W_\mathrm{s}$ values of interest with bonding characteristics of the M--X bond (M = Ba, Yb; X = C, O, F). Espinoza criterion $\frac{|V(\vv{r}_c)|}{G(\vv{r}_c)}$ is a measure of covalent bond character. $q_\mathrm{M}$ represents the natural charge of the heavy element.}
 \label{tab:comparison-bond}
\end{table}

\section{Experimental Considerations}

BaCH$_3$ and YbCH$_3$ are both prime candidates for creation of intense beams via buffer gas cooling \cite{Hutzler2012}.  The rotational spectrum of the ground electronic state of BaCH$_3$ has been studied via millimeter/submillimeter absorption~\cite{Xin1998} and was created via the reaction of barium metal with Sn(CH$_3$)$_4$.  Other alkaline-earth monomethyl molecules have been created via reaction of metals with other monomethyl species~\cite{Brazier1987}, including chloromethane (CH$_3$Cl) which could be used to react with ablated Yb or Ba metal in a cryogenic buffer gas cell.  While YbCH$_3$ has not been studied spectroscopically, the chemical similarity of Yb with the alkaline-earth metals means that these production methods would likely work as well, as has been demonstrated with a number of other Yb-containing molecules~\cite{Jadbabaie2020,augenbraun2021observation}.

The P, T-violation measurement with these species would be performed in the $K=1$ rotational state of the ground electronic and vibrational state, in which $K-$doubling splits the rotational states into doublets of opposite parity~\cite{kozyryev2017precision,Yu2021,augenbraun2021observation}.  The state with $K=1$ corresponds to the molecule rotating about its $C_{3v}$ symmetry axis with one quantum of angular momentum.  Alkaline-earth metals bound to $-$CH$_3$ and $-$OCH$_3$ ligands have this state at roughly 160~GHz above the ground state~\cite{Namiki1999,Sheridan2005,Yu2021}, giving them lifetimes estimated to be minutes or longer.  As this energy corresponds to $\approx 8$~K, there will be appreciable population in a buffer gas source at a few Kelvin.  The doubling in small, open-shell molecules with a metal-centered electron and a CH$_3$ group is dominated by anisotropic dipole-dipole interactions between the H nuclear spins and unpaired electron spin~\cite{Namiki1998,augenbraun2021observation}.  For CaOCH$_3$ this has been measured to be around 300 kHz~\cite{Namiki1998}, but due to the closer distance between the metal and H$_3$ group in MCH$_3$ molecules this value is likely a few times larger, around 1 MHz.  Note that the overall level structure, including the states which would be used for a spin precession measurement, are similar to other symmetric top molecules~\cite{Yu2021}.  Also similar is that the EDM sensitivity would saturate to $\approx 50\%$ of the maximum values at the small fields required to mix the parity doublets~\cite{Yu2021,petrov2022sensitivity}.

\medskip

\section{Conclusions}
In this work, we report the enhancement factors $W_\mathrm{d}$ and $W_\mathrm{s}$ needed for the interpretation of possible eEDM measurements  on the promising molecules BaCH$_3$ and YbCH$_3$. We carried out a systematic study to devise a computationally feasible scheme that provides accurate predictions along with realistic uncertainties. 

We showed that the scalar-relativistic level of theory in combination with the coupled cluster method and the ANO-RCC basis sets is a feasible and accurate approach for the optimisation of small polyatomic molecules. Correlating the valence and core-valence $(n-1)$ and $(n-2)$ electrons gives essentially the same results as correlating all electrons. The geometry obtained with this approach is in a very good agreement with the available experimental reports (for BaCH$_3$).

The values of $W_\mathrm{d}$ and $W_\mathrm{s}$ calculated in this work for both BaCH$_3$ and YbCH$_3$ are slightly larger than in other Ba- and Yb-containing molecules. 
The relation between
the increased covalent character of the heavy atom bond
and the size of the $W_\mathrm{d}$ and $W_\mathrm{s}$ factors provides important insight for the search for promising candidates for precision experiments.

Accurate calculations of properties needed for interpretation of precision measurements in polyatomic molecules is a challenging task and it is necessary to methodically evaluate the effects of the most relevant computational parameters. In this work, we provide such analysis for the BaCH$_3$ and YbCH$_3$ molecules. We show that calculating the $W_\mathrm{d}$ and $W_\mathrm{s}$ values requires an accurate description of the electron correlation. Specifically, in the coupled cluster approach, the number of electrons included in the correlation description plays the most significant role. The size of the basis set has a comparatively smaller but non-negligible effect. The systematic evaluation of the effect of various computational parameters also allowed us to estimate the uncertainty in the values presented in this work. 

\section*{Acknowledgements}
This publication is part of the project \textit{High Sector Fock space coupled cluster method: benchmark
accuracy across the periodic table}
(with project number VI.Vidi.192.088 of the research programme Vidi which is  financed by the Dutch Research Council (NWO)). We thank the Center for Information Technology of the University of Groningen for their support and for providing access to the Peregrine high performance computing cluster.
LFP acknowledges the support from the Slovak Research and Development Agency (APVV-20-0098, APVV-20-0127).

\providecommand{\noopsort}[1]{}\providecommand{\singleletter}[1]{#1}%
%

%
%


\section{Appendix}\label{sec:Appendix}

\subsection{Geometry optimization}\label{sec:Appendix-opt}

\subsubsection{Treatment of relativity}

Relativistic effects play an indispensable role in the correct description of molecules containing heavy elements. However, inclusion of just the scalar relativistic effects is often sufficient to describe properties that are not
very sensitive to the spin components, 
such as molecular geometry. This of course holds only for systems where the spin-orbit effects are not expected to be significant for the valence electrons, such as closed shell molecules or systems with valence $s$/$\sigma$ orbitals. We test this assumption by comparing the geometries of BaCH$_3$ obtained within the 4-component (4c) and the exact 2-component (X2C) \cite{iliavs2007infinite,saue2011relativistic} approaches and using the spin-free exact two-component one-electron variant formalism (SR)  \cite{zou2011development}. The first two methods were employed within the DIRAC19 program, while the SR  calculations were performed using the CFOUR package. In Table \ref{tab:opt-hamiltonian_basis}, it is observed that the X2C approach predicts the same results as the Dirac-Coulomb 4c Hamiltonian. Similarly the SR calculations differ in only 0.5\% to the 4c and X2C results. Consequently, the use of the computationally less expensive SR level of theory is justified in the geometry optimisation of BaCH$_3$ and YbCH$_3$ and similar molecules.

\subsubsection{Contracted vs. uncontracted basis set}

In electronic structure calculations, contracted basis functions constructed from Gaussians perform for many properties with similar accuracy but with dramatically reduced computational costs compared to uncontracted basis sets. In the geometry optimisation of BaCH$_3$ and YbCH$_3$ (Table~\ref{tab:opt-hamiltonian_basis}), we observe that the scalar-relativistic approximation introduces a smaller error compared to the effect of the basis set contraction. The difference between the results obtained using contracted ANO-RCC.VDZ and uncontracted dyall.v2z basis sets at the SR level is 1.7\%. However, this difference is reduced to 0.8\% when comparing the larger SR-ANO-RCC.VTZ and X2C-dyall.v3z methods. This is expected, since with the increasing cardinality, the basis sets become more saturated and thus the negative effect of contraction should decrease. We thus proceed with the computationally less expensive ANO-RCC basis sets. Based on the results in Table~\ref{tab:opt-hamiltonian_basis}, when selecting basis sets for geometry optimisations considering their cost-benefit values, it is advisable to choose a higher cardinality contracted basis set over an uncontracted but lower cardinality one.

\begin{table} 
\centering
\begin{tabular}{lllll}
    \hline\hline
Rel. & Basis set  & Ba--C [\AA] & C--H [\AA] & BaCH [$^\circ$] \\
    \hline
    4c & dyall.v2z   & 2.670 & 1.110 & 113.6 \\
    X2C & dyall.v2z & 2.670 & 1.110 & 113.6 \\
    SR  & dyall.v2z & 2.657 & 1.108 & 113.2\\
    SR  & ANO.VDZ & 2.613 & 1.107 & 112.7\\
    \hline
    X2C & dyall.v3z   & 2.605 & 1.100 & 112.8 \\
    SR  & ANO.VTZ & 2.585  & 1.098 & 112.6 \\
    \hline 
\end{tabular}
    \caption{Effect of the different levels of treatment of relativity on the optimised geometry of BaCH$_3$ at the CCSD level of theory. In all cases, 37 electrons were correlated.}
    \label{tab:opt-hamiltonian_basis}
\end{table}  

\subsubsection{Electron correlation}

Table \ref{tab:opt-triples} contains comparison between the results obtained within the CCSD and the CCSD(T) approaches. The calculations were carried out within the single-reference framework and 37 electrons were correlated. We observe that inclusion of perturbative triple excitations has a minor effect on the obtained geometry (0.2\% and 0.3\% at the DZ and TZ basis set cardinality, respectively) and thus conclude that optimisation on the CCSD level is of sufficient quality when computational resources are of importance. 

Furthermore, as expected, we find that the molecular geometry is a property mainly dependent on the valence region of the molecule. Table \ref{tab:opt-triples} shows that calculations correlating 37 electrons reproduce the all-electron results. 

\begin{table} 
\centering
    \begin{tabular}{lllll}
      \hline\hline
     Method  & $N$ & Ba--C [\AA] & C--H [\AA] & BaCH [$^\circ$] \\
    \hline
    &\multicolumn{4}{c}{ANO-RCC.VDZ}\\
    CCSD & 37 & 2.613 & 1.107 & 113 \\
    CCSD(T) & 37 & 2.609 & 1.109 &	113 \\
    \hline
    &\multicolumn{4}{c}{ANO-RCC.VTZ}\\
    CCSD & 37 & 2.586 & 1.098 & 113 \\
    CCSD(T) & 37 & 2.579 & 1.099 & 113 \\
    CCSD(T) & 65 & 2.578  & 1.099  & 113 \\ 
    CCSD(T) & 27 & 2.591  & 1.102  & 113  \\
    CCSD(T) & 17 & 2.606  & 1.102  & 113  \\
    \hline
    \end{tabular}
    \caption{Effect of the perturbative triple excitations on the optimised geometry of BaCH$_3$, at the SR level of theory. $N$ represents the number of correlated electrons.}
    \label{tab:opt-triples}
\end{table}

\subsection{Numerical stability of the FFPT}\label{sec:Appendix-ffpt}

\begin{figure}[ht!]
    \centering
    \includegraphics[scale=0.45]{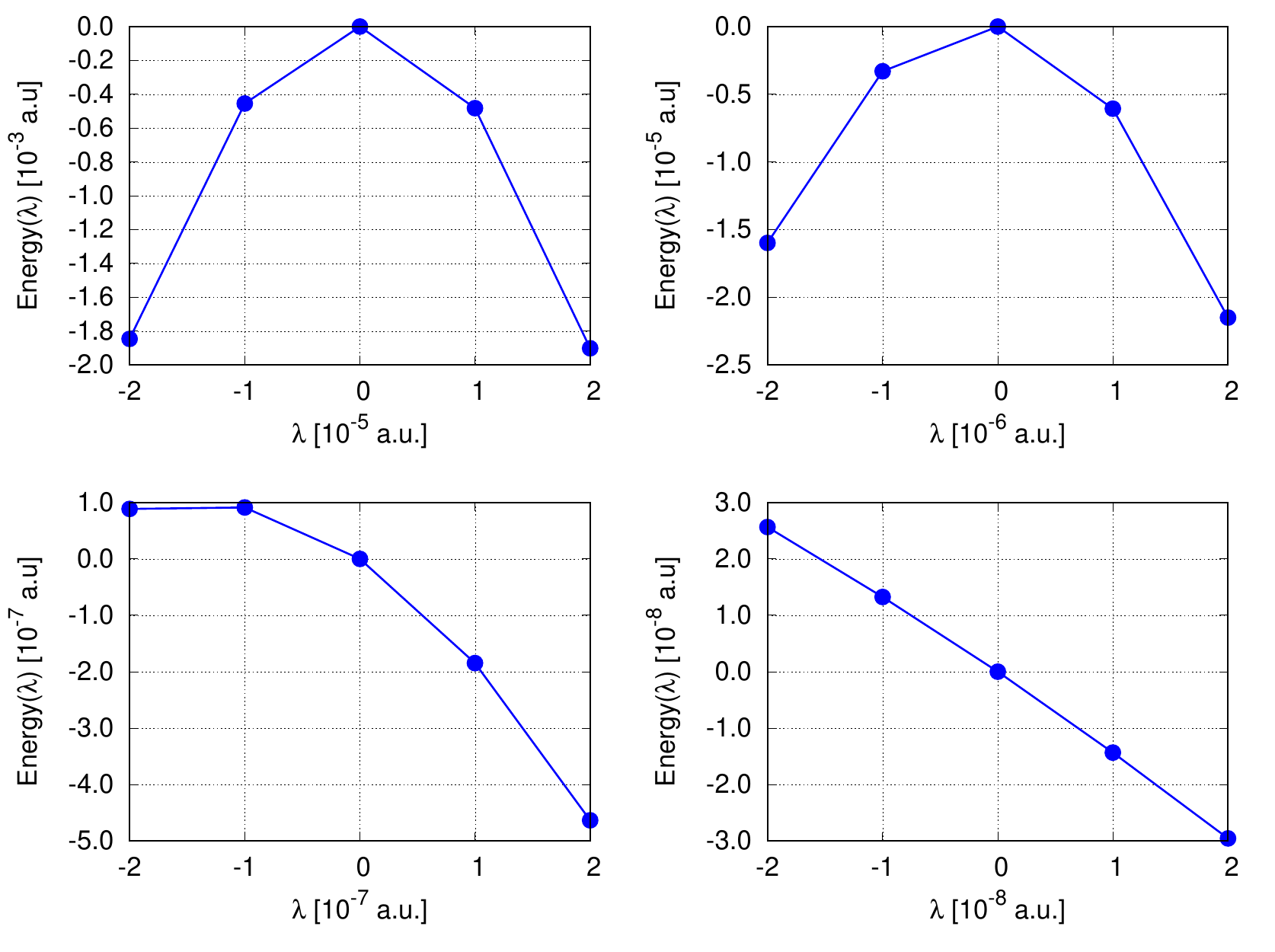}
    \caption{Energy dependency with the field strength in BaCH$_3$. 4c-CCSD and dyall.v2z basis set were used.}
    \label{fig:Wd-field}
\end{figure}

According to equations \eqref{eq:derivative1} and \eqref{eq:derivative2}, the $W_\mathrm{d}$ and $W_\mathrm{s}$  factors can be calculated as the first derivatives of the energy with respect to the corresponding perturbation.  To calculate the derivative using numerical differentiation, it is necessary to determine the field strength at which the total energy depends linearly on the perturbation. Figure~\ref{fig:Wd-field} shows the dependence of the total energy in BaCH$_3$ on the $\lambda_{d_\mathrm{e}}$ perturbation. Linear behavior is found at smaller fields of the order $\lambda_{d_\mathrm{e}}= 10^{-8}$~a.u. Similarly, for the $W_\mathrm{s}$ factor, the total energy is found to be linear at $\lambda_{k_\mathrm{s}}= 10^{-7}$~a.u. In YbCH$_3$, $\lambda_{d_\mathrm{e}}=\lambda_{k_\mathrm{s}}= 10^{-6}$~a.u. correspond to the linear regime. To support these small fields, the convergence criterion of the coupled cluster energy as well as the Hartree--Fock energy
was fixed at $10^{-11}$~a.u. 

\begin{thebibliography}{93}%
\makeatletter
\providecommand \@ifxundefined [1]{%
 \@ifx{#1\undefined}
}%
\providecommand \@ifnum [1]{%
 \ifnum #1\expandafter \@firstoftwo
 \else \expandafter \@secondoftwo
 \fi
}%
\providecommand \@ifx [1]{%
 \ifx #1\expandafter \@firstoftwo
 \else \expandafter \@secondoftwo
 \fi
}%
\providecommand \natexlab [1]{#1}%
\providecommand \enquote  [1]{``#1''}%
\providecommand \bibnamefont  [1]{#1}%
\providecommand \bibfnamefont [1]{#1}%
\providecommand \citenamefont [1]{#1}%
\providecommand \href@noop [0]{\@secondoftwo}%
\providecommand \href [0]{\begingroup \@sanitize@url \@href}%
\providecommand \@href[1]{\@@startlink{#1}\@@href}%
\providecommand \@@href[1]{\endgroup#1\@@endlink}%
\providecommand \@sanitize@url [0]{\catcode `\\12\catcode `\$12\catcode
  `\&12\catcode `\#12\catcode `\^12\catcode `\_12\catcode `\%12\relax}%
\providecommand \@@startlink[1]{}%
\providecommand \@@endlink[0]{}%
\providecommand \url  [0]{\begingroup\@sanitize@url \@url }%
\providecommand \@url [1]{\endgroup\@href {#1}{\urlprefix }}%
\providecommand \urlprefix  [0]{URL }%
\providecommand \Eprint [0]{\href }%
\providecommand \doibase [0]{https://doi.org/}%
\providecommand \selectlanguage [0]{\@gobble}%
\providecommand \bibinfo  [0]{\@secondoftwo}%
\providecommand \bibfield  [0]{\@secondoftwo}%
\providecommand \translation [1]{[#1]}%
\providecommand \BibitemOpen [0]{}%
\providecommand \bibitemStop [0]{}%
\providecommand \bibitemNoStop [0]{.\EOS\space}%
\providecommand \EOS [0]{\spacefactor3000\relax}%
\providecommand \BibitemShut  [1]{\csname bibitem#1\endcsname}%
\let\auto@bib@innerbib\@empty
\bibitem [{\citenamefont {Bertone}(2010)}]{bertone2010particle}%
  \BibitemOpen
  \bibfield  {author} {\bibinfo {author} {\bibfnamefont {G.}~\bibnamefont
  {Bertone}},\ }\href@noop {} {\emph {\bibinfo {title} {Particle dark matter:
  observations, models and searches}}}\ (\bibinfo  {publisher} {Cambridge
  University Press},\ \bibinfo {year} {2010})\BibitemShut {NoStop}%
\bibitem [{\citenamefont {Solà~Peracaula}(2013)}]{sola2013cosmological}%
  \BibitemOpen
  \bibfield  {author} {\bibinfo {author} {\bibfnamefont {J.}~\bibnamefont
  {Solà~Peracaula}},\ }\bibfield  {title} {\bibinfo {title} {Cosmological
  constant and vacuum energy: Old and new ideas},\ }\href
  {https://doi.org/10.1088/1742-6596/453/1/012015} {\bibfield  {journal}
  {\bibinfo  {journal} {Journal of Physics Conference Series}\ }\textbf
  {\bibinfo {volume} {453}} (\bibinfo {year} {2013})}\BibitemShut {NoStop}%
\bibitem [{\citenamefont {Dine}\ and\ \citenamefont
  {Kusenko}(2003)}]{dine2003origin}%
  \BibitemOpen
  \bibfield  {author} {\bibinfo {author} {\bibfnamefont {M.}~\bibnamefont
  {Dine}}\ and\ \bibinfo {author} {\bibfnamefont {A.}~\bibnamefont {Kusenko}},\
  }\bibfield  {title} {\bibinfo {title} {Origin of the matter-antimatter
  asymmetry},\ }\href@noop {} {\bibfield  {journal} {\bibinfo  {journal}
  {Reviews of Modern Physics}\ }\textbf {\bibinfo {volume} {76}},\ \bibinfo
  {pages} {1} (\bibinfo {year} {2003})}\BibitemShut {NoStop}%
\bibitem [{\citenamefont {Ginges}\ and\ \citenamefont
  {Flambaum}(2004)}]{ginges2004violations}%
  \BibitemOpen
  \bibfield  {author} {\bibinfo {author} {\bibfnamefont {J.}~\bibnamefont
  {Ginges}}\ and\ \bibinfo {author} {\bibfnamefont {V.~V.}\ \bibnamefont
  {Flambaum}},\ }\bibfield  {title} {\bibinfo {title} {Violations of
  fundamental symmetries in atoms and tests of unification theories of
  elementary particles},\ }\href@noop {} {\bibfield  {journal} {\bibinfo
  {journal} {Physics Reports}\ }\textbf {\bibinfo {volume} {397}},\ \bibinfo
  {pages} {63} (\bibinfo {year} {2004})}\BibitemShut {NoStop}%
\bibitem [{\citenamefont {Yamaguchi}\ and\ \citenamefont
  {Yamanaka}(2020)}]{yamaguchi2020large}%
  \BibitemOpen
  \bibfield  {author} {\bibinfo {author} {\bibfnamefont {Y.}~\bibnamefont
  {Yamaguchi}}\ and\ \bibinfo {author} {\bibfnamefont {N.}~\bibnamefont
  {Yamanaka}},\ }\bibfield  {title} {\bibinfo {title} {Large long-distance
  contributions to the electric dipole moments of charged leptons in the
  standard model},\ }\href@noop {} {\bibfield  {journal} {\bibinfo  {journal}
  {Physical Review Letters}\ }\textbf {\bibinfo {volume} {125}},\ \bibinfo
  {pages} {241802} (\bibinfo {year} {2020})}\BibitemShut {NoStop}%
\bibitem [{\citenamefont {Commins}(1999)}]{commins1999electric}%
  \BibitemOpen
  \bibfield  {author} {\bibinfo {author} {\bibfnamefont {E.~D.}\ \bibnamefont
  {Commins}},\ }\bibfield  {title} {\bibinfo {title} {Electric dipole moments
  of leptons},\ }in\ \href@noop {} {\emph {\bibinfo {booktitle} {Advances in
  Atomic, Molecular, and Optical Physics}}},\ Vol.~\bibinfo {volume} {40}\
  (\bibinfo  {publisher} {Elsevier},\ \bibinfo {year} {1999})\ pp.\ \bibinfo
  {pages} {1--55}\BibitemShut {NoStop}%
\bibitem [{\citenamefont {Bernreuther}\ and\ \citenamefont
  {Suzuki}(1991)}]{bernreuther1991electric}%
  \BibitemOpen
  \bibfield  {author} {\bibinfo {author} {\bibfnamefont {W.}~\bibnamefont
  {Bernreuther}}\ and\ \bibinfo {author} {\bibfnamefont {M.}~\bibnamefont
  {Suzuki}},\ }\bibfield  {title} {\bibinfo {title} {The electric dipole moment
  of the electron},\ }\href@noop {} {\bibfield  {journal} {\bibinfo  {journal}
  {Reviews of Modern Physics}\ }\textbf {\bibinfo {volume} {63}},\ \bibinfo
  {pages} {313} (\bibinfo {year} {1991})}\BibitemShut {NoStop}%
\bibitem [{\citenamefont {Pospelov}\ and\ \citenamefont
  {Ritz}(2014)}]{pospelov2014ckm}%
  \BibitemOpen
  \bibfield  {author} {\bibinfo {author} {\bibfnamefont {M.}~\bibnamefont
  {Pospelov}}\ and\ \bibinfo {author} {\bibfnamefont {A.}~\bibnamefont
  {Ritz}},\ }\bibfield  {title} {\bibinfo {title} {{CKM} benchmarks for
  electron electric dipole moment experiments},\ }\href@noop {} {\bibfield
  {journal} {\bibinfo  {journal} {Physical Review D}\ }\textbf {\bibinfo
  {volume} {89}},\ \bibinfo {pages} {056006} (\bibinfo {year}
  {2014})}\BibitemShut {NoStop}%
\bibitem [{\citenamefont {Sachs}\ and\ \citenamefont
  {Schwebel}(1959)}]{sachs1959implications}%
  \BibitemOpen
  \bibfield  {author} {\bibinfo {author} {\bibfnamefont {M.}~\bibnamefont
  {Sachs}}\ and\ \bibinfo {author} {\bibfnamefont {S.~L.}\ \bibnamefont
  {Schwebel}},\ }\bibfield  {title} {\bibinfo {title} {Implications of parity
  nonconservation and time reversal noninvariance in electromagnetic
  interactions: {P}art {II}. {A}tomic energy levels},\ }\href@noop {}
  {\bibfield  {journal} {\bibinfo  {journal} {Annals of Physics}\ }\textbf
  {\bibinfo {volume} {8}},\ \bibinfo {pages} {475} (\bibinfo {year}
  {1959})}\BibitemShut {NoStop}%
\bibitem [{\citenamefont {Salpeter}(1958)}]{salpeter1958some}%
  \BibitemOpen
  \bibfield  {author} {\bibinfo {author} {\bibfnamefont {E.}~\bibnamefont
  {Salpeter}},\ }\bibfield  {title} {\bibinfo {title} {Some atomic effects of
  an electronic electric dipole moment},\ }\href@noop {} {\bibfield  {journal}
  {\bibinfo  {journal} {Physical Review}\ }\textbf {\bibinfo {volume} {112}},\
  \bibinfo {pages} {1642} (\bibinfo {year} {1958})}\BibitemShut {NoStop}%
\bibitem [{\citenamefont {Bouchiat}\ and\ \citenamefont
  {Bouchiat}(1974)}]{bouchiat1974parity}%
  \BibitemOpen
  \bibfield  {author} {\bibinfo {author} {\bibfnamefont {M.}~\bibnamefont
  {Bouchiat}}\ and\ \bibinfo {author} {\bibfnamefont {C.}~\bibnamefont
  {Bouchiat}},\ }\bibfield  {title} {\bibinfo {title} {{I}. {P}arity violation
  induced by weak neutral currents in atomic physics},\ }\href@noop {}
  {\bibfield  {journal} {\bibinfo  {journal} {Journal de Physique}\ }\textbf
  {\bibinfo {volume} {35}},\ \bibinfo {pages} {899} (\bibinfo {year}
  {1974})}\BibitemShut {NoStop}%
\bibitem [{\citenamefont {Sandars}(1965)}]{sandars1965electric}%
  \BibitemOpen
  \bibfield  {author} {\bibinfo {author} {\bibfnamefont {P.}~\bibnamefont
  {Sandars}},\ }\bibfield  {title} {\bibinfo {title} {The electric dipole
  moment of an atom},\ }\href@noop {} {\bibfield  {journal} {\bibinfo
  {journal} {Physics Letters}\ }\textbf {\bibinfo {volume} {14}},\ \bibinfo
  {pages} {194} (\bibinfo {year} {1965})}\BibitemShut {NoStop}%
\bibitem [{\citenamefont {Sandars}(1968)}]{sandars1968electric}%
  \BibitemOpen
  \bibfield  {author} {\bibinfo {author} {\bibfnamefont {P.}~\bibnamefont
  {Sandars}},\ }\bibfield  {title} {\bibinfo {title} {The electric-dipole
  moments of an atom {II}. the contribution from an electric-dipole moment on
  the electron with particular reference to the hydrogen atom},\ }\href@noop {}
  {\bibfield  {journal} {\bibinfo  {journal} {Journal of Physics B: Atomic and
  Molecular Physics}\ }\textbf {\bibinfo {volume} {1}},\ \bibinfo {pages} {511}
  (\bibinfo {year} {1968})}\BibitemShut {NoStop}%
\bibitem [{\citenamefont {Sandars}(1967)}]{sandars1967measurability}%
  \BibitemOpen
  \bibfield  {author} {\bibinfo {author} {\bibfnamefont {P.}~\bibnamefont
  {Sandars}},\ }\bibfield  {title} {\bibinfo {title} {Measurability of the
  proton electric dipole moment},\ }\href@noop {} {\bibfield  {journal}
  {\bibinfo  {journal} {Physical Review Letters}\ }\textbf {\bibinfo {volume}
  {19}},\ \bibinfo {pages} {1396} (\bibinfo {year} {1967})}\BibitemShut
  {NoStop}%
\bibitem [{\citenamefont {DeMille}(2015)}]{demille2015diatomic}%
  \BibitemOpen
  \bibfield  {author} {\bibinfo {author} {\bibfnamefont {D.}~\bibnamefont
  {DeMille}},\ }\bibfield  {title} {\bibinfo {title} {Diatomic molecules, a
  window onto fundamental physics},\ }\href@noop {} {\bibfield  {journal}
  {\bibinfo  {journal} {Physics Today}\ }\textbf {\bibinfo {volume} {68}},\
  \bibinfo {pages} {34} (\bibinfo {year} {2015})}\BibitemShut {NoStop}%
\bibitem [{\citenamefont {Harrison}\ \emph {et~al.}(1969)\citenamefont
  {Harrison}, \citenamefont {Sandars},\ and\ \citenamefont
  {Wright}}]{harrison1969experimental}%
  \BibitemOpen
  \bibfield  {author} {\bibinfo {author} {\bibfnamefont {G.}~\bibnamefont
  {Harrison}}, \bibinfo {author} {\bibfnamefont {P.}~\bibnamefont {Sandars}},\
  and\ \bibinfo {author} {\bibfnamefont {S.}~\bibnamefont {Wright}},\
  }\bibfield  {title} {\bibinfo {title} {Experimental limit on the proton
  electric dipole moment},\ }\href@noop {} {\bibfield  {journal} {\bibinfo
  {journal} {Physical Review Letters}\ }\textbf {\bibinfo {volume} {22}},\
  \bibinfo {pages} {1263} (\bibinfo {year} {1969})}\BibitemShut {NoStop}%
\bibitem [{\citenamefont {Hinds}\ \emph {et~al.}(1976)\citenamefont {Hinds},
  \citenamefont {Loving},\ and\ \citenamefont {Sandars}}]{hinds1976limits}%
  \BibitemOpen
  \bibfield  {author} {\bibinfo {author} {\bibfnamefont {E.}~\bibnamefont
  {Hinds}}, \bibinfo {author} {\bibfnamefont {C.}~\bibnamefont {Loving}},\ and\
  \bibinfo {author} {\bibfnamefont {P.}~\bibnamefont {Sandars}},\ }\bibfield
  {title} {\bibinfo {title} {Limits on {P} and {T} violating neutral current
  weak interactions},\ }\href@noop {} {\bibfield  {journal} {\bibinfo
  {journal} {Physics Letters B}\ }\textbf {\bibinfo {volume} {62}},\ \bibinfo
  {pages} {97} (\bibinfo {year} {1976})}\BibitemShut {NoStop}%
\bibitem [{\citenamefont {Hinds}\ and\ \citenamefont
  {Sandars}(1980)}]{hinds1980experiment}%
  \BibitemOpen
  \bibfield  {author} {\bibinfo {author} {\bibfnamefont {E.~A.}\ \bibnamefont
  {Hinds}}\ and\ \bibinfo {author} {\bibfnamefont {P.}~\bibnamefont
  {Sandars}},\ }\bibfield  {title} {\bibinfo {title} {Experiment to search for
  {P}-and {T}-violating interactions in the hyperfine structure of thallium
  fluoride},\ }\href@noop {} {\bibfield  {journal} {\bibinfo  {journal}
  {Physical Review A}\ }\textbf {\bibinfo {volume} {21}},\ \bibinfo {pages}
  {480} (\bibinfo {year} {1980})}\BibitemShut {NoStop}%
\bibitem [{\citenamefont {Andreev}\ and\ \citenamefont
  {Hutzler}(2018)}]{andreev2018improved}%
  \BibitemOpen
  \bibfield  {author} {\bibinfo {author} {\bibfnamefont {V.}~\bibnamefont
  {Andreev}}\ and\ \bibinfo {author} {\bibfnamefont {N.}~\bibnamefont
  {Hutzler}},\ }\bibfield  {title} {\bibinfo {title} {Improved limit on the
  electric dipole moment of the electron},\ }\href@noop {} {\bibfield
  {journal} {\bibinfo  {journal} {Nature}\ }\textbf {\bibinfo {volume} {562}},\
  \bibinfo {pages} {355} (\bibinfo {year} {2018})}\BibitemShut {NoStop}%
\bibitem [{\citenamefont {Kara}\ \emph {et~al.}(2012)\citenamefont {Kara},
  \citenamefont {Smallman}, \citenamefont {Hudson}, \citenamefont {Sauer},
  \citenamefont {Tarbutt},\ and\ \citenamefont {Hinds}}]{kara2012measurement}%
  \BibitemOpen
  \bibfield  {author} {\bibinfo {author} {\bibfnamefont {D.~M.}\ \bibnamefont
  {Kara}}, \bibinfo {author} {\bibfnamefont {I.}~\bibnamefont {Smallman}},
  \bibinfo {author} {\bibfnamefont {J.~J.}\ \bibnamefont {Hudson}}, \bibinfo
  {author} {\bibfnamefont {B.~E.}\ \bibnamefont {Sauer}}, \bibinfo {author}
  {\bibfnamefont {M.~R.}\ \bibnamefont {Tarbutt}},\ and\ \bibinfo {author}
  {\bibfnamefont {E.~A.}\ \bibnamefont {Hinds}},\ }\bibfield  {title} {\bibinfo
  {title} {Measurement of the electron's electric dipole moment using {YbF}
  molecules: methods and data analysis},\ }\href@noop {} {\bibfield  {journal}
  {\bibinfo  {journal} {New Journal of Physics}\ }\textbf {\bibinfo {volume}
  {14}},\ \bibinfo {pages} {103051} (\bibinfo {year} {2012})}\BibitemShut
  {NoStop}%
\bibitem [{\citenamefont {Cairncross}\ \emph {et~al.}(2017)\citenamefont
  {Cairncross}, \citenamefont {Gresh}, \citenamefont {Grau}, \citenamefont
  {Cossel}, \citenamefont {Roussy}, \citenamefont {Ni}, \citenamefont {Zhou},
  \citenamefont {Ye},\ and\ \citenamefont {Cornell}}]{cairncross2017precision}%
  \BibitemOpen
  \bibfield  {author} {\bibinfo {author} {\bibfnamefont {W.~B.}\ \bibnamefont
  {Cairncross}}, \bibinfo {author} {\bibfnamefont {D.~N.}\ \bibnamefont
  {Gresh}}, \bibinfo {author} {\bibfnamefont {M.}~\bibnamefont {Grau}},
  \bibinfo {author} {\bibfnamefont {K.~C.}\ \bibnamefont {Cossel}}, \bibinfo
  {author} {\bibfnamefont {T.~S.}\ \bibnamefont {Roussy}}, \bibinfo {author}
  {\bibfnamefont {Y.}~\bibnamefont {Ni}}, \bibinfo {author} {\bibfnamefont
  {Y.}~\bibnamefont {Zhou}}, \bibinfo {author} {\bibfnamefont {J.}~\bibnamefont
  {Ye}},\ and\ \bibinfo {author} {\bibfnamefont {E.~A.}\ \bibnamefont
  {Cornell}},\ }\bibfield  {title} {\bibinfo {title} {Precision measurement of
  the electron’s electric dipole moment using trapped molecular ions},\
  }\href@noop {} {\bibfield  {journal} {\bibinfo  {journal} {Physical Review
  Letters}\ }\textbf {\bibinfo {volume} {119}},\ \bibinfo {pages} {153001}
  (\bibinfo {year} {2017})}\BibitemShut {NoStop}%
\bibitem [{\citenamefont {Garcia~Ruiz}\ \emph {et~al.}(2020)\citenamefont
  {Garcia~Ruiz}, \citenamefont {Berger}, \citenamefont {Billowes},
  \citenamefont {Binnersley}, \citenamefont {Bissell}, \citenamefont {Breier},
  \citenamefont {Brinson}, \citenamefont {Chrysalidis}, \citenamefont
  {Cocolios}, \citenamefont {Cooper} \emph {et~al.}}]{garcia2020spectroscopy}%
  \BibitemOpen
  \bibfield  {author} {\bibinfo {author} {\bibfnamefont {R.~F.}\ \bibnamefont
  {Garcia~Ruiz}}, \bibinfo {author} {\bibfnamefont {R.}~\bibnamefont {Berger}},
  \bibinfo {author} {\bibfnamefont {J.}~\bibnamefont {Billowes}}, \bibinfo
  {author} {\bibfnamefont {C.}~\bibnamefont {Binnersley}}, \bibinfo {author}
  {\bibfnamefont {M.}~\bibnamefont {Bissell}}, \bibinfo {author} {\bibfnamefont
  {A.}~\bibnamefont {Breier}}, \bibinfo {author} {\bibfnamefont
  {A.}~\bibnamefont {Brinson}}, \bibinfo {author} {\bibfnamefont
  {K.}~\bibnamefont {Chrysalidis}}, \bibinfo {author} {\bibfnamefont
  {T.}~\bibnamefont {Cocolios}}, \bibinfo {author} {\bibfnamefont
  {B.}~\bibnamefont {Cooper}}, \emph {et~al.},\ }\bibfield  {title} {\bibinfo
  {title} {Spectroscopy of short-lived radioactive molecules},\ }\href@noop {}
  {\bibfield  {journal} {\bibinfo  {journal} {Nature}\ }\textbf {\bibinfo
  {volume} {581}},\ \bibinfo {pages} {396} (\bibinfo {year}
  {2020})}\BibitemShut {NoStop}%
\bibitem [{\citenamefont {Isaev}\ \emph {et~al.}(2010)\citenamefont {Isaev},
  \citenamefont {Hoekstra},\ and\ \citenamefont {Berger}}]{isaev2010laser}%
  \BibitemOpen
  \bibfield  {author} {\bibinfo {author} {\bibfnamefont {T.}~\bibnamefont
  {Isaev}}, \bibinfo {author} {\bibfnamefont {S.}~\bibnamefont {Hoekstra}},\
  and\ \bibinfo {author} {\bibfnamefont {R.}~\bibnamefont {Berger}},\
  }\bibfield  {title} {\bibinfo {title} {Laser-cooled {RaF} as a promising
  candidate to measure molecular parity violation},\ }\href@noop {} {\bibfield
  {journal} {\bibinfo  {journal} {Physical Review A}\ }\textbf {\bibinfo
  {volume} {82}},\ \bibinfo {pages} {052521} (\bibinfo {year}
  {2010})}\BibitemShut {NoStop}%
\bibitem [{\citenamefont {Aggarwal}\ \emph {et~al.}(2018)\citenamefont
  {Aggarwal}, \citenamefont {Bethlem}, \citenamefont {Borschevsky},
  \citenamefont {Denis}, \citenamefont {Esajas}, \citenamefont {Haase},
  \citenamefont {Hao}, \citenamefont {Hoekstra}, \citenamefont {Jungmann},
  \citenamefont {Meijknecht} \emph {et~al.}}]{aggarwal2018measuring}%
  \BibitemOpen
  \bibfield  {author} {\bibinfo {author} {\bibfnamefont {P.}~\bibnamefont
  {Aggarwal}}, \bibinfo {author} {\bibfnamefont {H.~L.}\ \bibnamefont
  {Bethlem}}, \bibinfo {author} {\bibfnamefont {A.}~\bibnamefont
  {Borschevsky}}, \bibinfo {author} {\bibfnamefont {M.}~\bibnamefont {Denis}},
  \bibinfo {author} {\bibfnamefont {K.}~\bibnamefont {Esajas}}, \bibinfo
  {author} {\bibfnamefont {P.~A.}\ \bibnamefont {Haase}}, \bibinfo {author}
  {\bibfnamefont {Y.}~\bibnamefont {Hao}}, \bibinfo {author} {\bibfnamefont
  {S.}~\bibnamefont {Hoekstra}}, \bibinfo {author} {\bibfnamefont
  {K.}~\bibnamefont {Jungmann}}, \bibinfo {author} {\bibfnamefont {T.~B.}\
  \bibnamefont {Meijknecht}}, \emph {et~al.},\ }\bibfield  {title} {\bibinfo
  {title} {Measuring the electric dipole moment of the electron in {BaF}},\
  }\href@noop {} {\bibfield  {journal} {\bibinfo  {journal} {The European
  Physical Journal D}\ }\textbf {\bibinfo {volume} {72}},\ \bibinfo {pages} {1}
  (\bibinfo {year} {2018})}\BibitemShut {NoStop}%
\bibitem [{\citenamefont {Bickman}\ \emph {et~al.}(2009)\citenamefont
  {Bickman}, \citenamefont {Hamilton}, \citenamefont {Jiang},\ and\
  \citenamefont {DeMille}}]{bickman2009preparation}%
  \BibitemOpen
  \bibfield  {author} {\bibinfo {author} {\bibfnamefont {S.}~\bibnamefont
  {Bickman}}, \bibinfo {author} {\bibfnamefont {P.}~\bibnamefont {Hamilton}},
  \bibinfo {author} {\bibfnamefont {Y.}~\bibnamefont {Jiang}},\ and\ \bibinfo
  {author} {\bibfnamefont {D.}~\bibnamefont {DeMille}},\ }\bibfield  {title}
  {\bibinfo {title} {Preparation and detection of states with simultaneous spin
  alignment and selectable molecular orientation in {PbO}},\ }\href@noop {}
  {\bibfield  {journal} {\bibinfo  {journal} {Physical Review A}\ }\textbf
  {\bibinfo {volume} {80}},\ \bibinfo {pages} {023418} (\bibinfo {year}
  {2009})}\BibitemShut {NoStop}%
\bibitem [{\citenamefont {Hudson}\ \emph {et~al.}(2011)\citenamefont {Hudson},
  \citenamefont {Kara}, \citenamefont {Smallman}, \citenamefont {Sauer},
  \citenamefont {Tarbutt},\ and\ \citenamefont {Hinds}}]{hudson2011improved}%
  \BibitemOpen
  \bibfield  {author} {\bibinfo {author} {\bibfnamefont {J.~J.}\ \bibnamefont
  {Hudson}}, \bibinfo {author} {\bibfnamefont {D.~M.}\ \bibnamefont {Kara}},
  \bibinfo {author} {\bibfnamefont {I.}~\bibnamefont {Smallman}}, \bibinfo
  {author} {\bibfnamefont {B.~E.}\ \bibnamefont {Sauer}}, \bibinfo {author}
  {\bibfnamefont {M.~R.}\ \bibnamefont {Tarbutt}},\ and\ \bibinfo {author}
  {\bibfnamefont {E.~A.}\ \bibnamefont {Hinds}},\ }\bibfield  {title} {\bibinfo
  {title} {Improved measurement of the shape of the electron},\ }\href@noop {}
  {\bibfield  {journal} {\bibinfo  {journal} {Nature}\ }\textbf {\bibinfo
  {volume} {473}},\ \bibinfo {pages} {493} (\bibinfo {year}
  {2011})}\BibitemShut {NoStop}%
\bibitem [{\citenamefont {Kozyryev}\ and\ \citenamefont
  {Hutzler}(2017)}]{kozyryev2017precision}%
  \BibitemOpen
  \bibfield  {author} {\bibinfo {author} {\bibfnamefont {I.}~\bibnamefont
  {Kozyryev}}\ and\ \bibinfo {author} {\bibfnamefont {N.~R.}\ \bibnamefont
  {Hutzler}},\ }\bibfield  {title} {\bibinfo {title} {Precision measurement of
  time-reversal symmetry violation with laser-cooled polyatomic molecules},\
  }\href@noop {} {\bibfield  {journal} {\bibinfo  {journal} {Physical Review
  Letters}\ }\textbf {\bibinfo {volume} {119}},\ \bibinfo {pages} {133002}
  (\bibinfo {year} {2017})}\BibitemShut {NoStop}%
\bibitem [{\citenamefont {Hutzler}(2020)}]{hutzler2020polyatomic}%
  \BibitemOpen
  \bibfield  {author} {\bibinfo {author} {\bibfnamefont {N.~R.}\ \bibnamefont
  {Hutzler}},\ }\bibfield  {title} {\bibinfo {title} {Polyatomic molecules as
  quantum sensors for fundamental physics},\ }\href@noop {} {\bibfield
  {journal} {\bibinfo  {journal} {Quantum Science and Technology}\ }\textbf
  {\bibinfo {volume} {5}},\ \bibinfo {pages} {044011} (\bibinfo {year}
  {2020})}\BibitemShut {NoStop}%
\bibitem [{\citenamefont {Ellis}(2001)}]{ellis2001main}%
  \BibitemOpen
  \bibfield  {author} {\bibinfo {author} {\bibfnamefont {A.~M.}\ \bibnamefont
  {Ellis}},\ }\bibfield  {title} {\bibinfo {title} {Main group metal-ligand
  interactions in small molecules: new insights from laser spectroscopy},\
  }\href@noop {} {\bibfield  {journal} {\bibinfo  {journal} {International
  Reviews in Physical Chemistry}\ }\textbf {\bibinfo {volume} {20}},\ \bibinfo
  {pages} {551} (\bibinfo {year} {2001})}\BibitemShut {NoStop}%
\bibitem [{\citenamefont {Isaev}\ \emph {et~al.}(2017)\citenamefont {Isaev},
  \citenamefont {Zaitsevskii},\ and\ \citenamefont {Eliav}}]{isaev2017laser}%
  \BibitemOpen
  \bibfield  {author} {\bibinfo {author} {\bibfnamefont {T.}~\bibnamefont
  {Isaev}}, \bibinfo {author} {\bibfnamefont {A.}~\bibnamefont {Zaitsevskii}},\
  and\ \bibinfo {author} {\bibfnamefont {E.}~\bibnamefont {Eliav}},\ }\bibfield
   {title} {\bibinfo {title} {Laser-coolable polyatomic molecules with heavy
  nuclei},\ }\href@noop {} {\bibfield  {journal} {\bibinfo  {journal} {Journal
  of Physics B: Atomic, Molecular and Optical Physics}\ }\textbf {\bibinfo
  {volume} {50}},\ \bibinfo {pages} {225101} (\bibinfo {year}
  {2017})}\BibitemShut {NoStop}%
\bibitem [{\citenamefont {Isaev}\ and\ \citenamefont
  {Berger}(2016)}]{isaev2016polyatomic}%
  \BibitemOpen
  \bibfield  {author} {\bibinfo {author} {\bibfnamefont {T.~A.}\ \bibnamefont
  {Isaev}}\ and\ \bibinfo {author} {\bibfnamefont {R.}~\bibnamefont {Berger}},\
  }\bibfield  {title} {\bibinfo {title} {Polyatomic candidates for cooling of
  molecules with lasers from simple theoretical concepts},\ }\href@noop {}
  {\bibfield  {journal} {\bibinfo  {journal} {Physical Review Letters}\
  }\textbf {\bibinfo {volume} {116}},\ \bibinfo {pages} {063006} (\bibinfo
  {year} {2016})}\BibitemShut {NoStop}%
\bibitem [{\citenamefont {Kozyryev}\ \emph {et~al.}(2016)\citenamefont
  {Kozyryev}, \citenamefont {Baum}, \citenamefont {Matsuda},\ and\
  \citenamefont {Doyle}}]{kozyryev2016proposal}%
  \BibitemOpen
  \bibfield  {author} {\bibinfo {author} {\bibfnamefont {I.}~\bibnamefont
  {Kozyryev}}, \bibinfo {author} {\bibfnamefont {L.}~\bibnamefont {Baum}},
  \bibinfo {author} {\bibfnamefont {K.}~\bibnamefont {Matsuda}},\ and\ \bibinfo
  {author} {\bibfnamefont {J.~M.}\ \bibnamefont {Doyle}},\ }\bibfield  {title}
  {\bibinfo {title} {Proposal for laser cooling of complex polyatomic
  molecules},\ }\href@noop {} {\bibfield  {journal} {\bibinfo  {journal}
  {ChemPhysChem}\ }\textbf {\bibinfo {volume} {17}},\ \bibinfo {pages} {3641}
  (\bibinfo {year} {2016})}\BibitemShut {NoStop}%
\bibitem [{\citenamefont {Mitra}\ \emph {et~al.}(2020)\citenamefont {Mitra},
  \citenamefont {Vilas}, \citenamefont {Hallas}, \citenamefont {Anderegg},
  \citenamefont {Augenbraun}, \citenamefont {Baum}, \citenamefont {Miller},
  \citenamefont {Raval},\ and\ \citenamefont {Doyle}}]{mitra2020direct}%
  \BibitemOpen
  \bibfield  {author} {\bibinfo {author} {\bibfnamefont {D.}~\bibnamefont
  {Mitra}}, \bibinfo {author} {\bibfnamefont {N.~B.}\ \bibnamefont {Vilas}},
  \bibinfo {author} {\bibfnamefont {C.}~\bibnamefont {Hallas}}, \bibinfo
  {author} {\bibfnamefont {L.}~\bibnamefont {Anderegg}}, \bibinfo {author}
  {\bibfnamefont {B.~L.}\ \bibnamefont {Augenbraun}}, \bibinfo {author}
  {\bibfnamefont {L.}~\bibnamefont {Baum}}, \bibinfo {author} {\bibfnamefont
  {C.}~\bibnamefont {Miller}}, \bibinfo {author} {\bibfnamefont
  {S.}~\bibnamefont {Raval}},\ and\ \bibinfo {author} {\bibfnamefont {J.~M.}\
  \bibnamefont {Doyle}},\ }\bibfield  {title} {\bibinfo {title} {Direct laser
  cooling of a symmetric top molecule},\ }\href@noop {} {\bibfield  {journal}
  {\bibinfo  {journal} {Science}\ }\textbf {\bibinfo {volume} {369}},\ \bibinfo
  {pages} {1366} (\bibinfo {year} {2020})}\BibitemShut {NoStop}%
\bibitem [{\citenamefont {Augenbraun}\ \emph {et~al.}(2021)\citenamefont
  {Augenbraun}, \citenamefont {Lasner}, \citenamefont {Frenett}, \citenamefont
  {Sawaoka}, \citenamefont {Le}, \citenamefont {Doyle},\ and\ \citenamefont
  {Steimle}}]{augenbraun2021observation}%
  \BibitemOpen
  \bibfield  {author} {\bibinfo {author} {\bibfnamefont {B.~L.}\ \bibnamefont
  {Augenbraun}}, \bibinfo {author} {\bibfnamefont {Z.~D.}\ \bibnamefont
  {Lasner}}, \bibinfo {author} {\bibfnamefont {A.}~\bibnamefont {Frenett}},
  \bibinfo {author} {\bibfnamefont {H.}~\bibnamefont {Sawaoka}}, \bibinfo
  {author} {\bibfnamefont {A.~T.}\ \bibnamefont {Le}}, \bibinfo {author}
  {\bibfnamefont {J.~M.}\ \bibnamefont {Doyle}},\ and\ \bibinfo {author}
  {\bibfnamefont {T.~C.}\ \bibnamefont {Steimle}},\ }\bibfield  {title}
  {\bibinfo {title} {Observation and laser spectroscopy of ytterbium
  monomethoxide, {YbOCH$_3$}},\ }\href@noop {} {\bibfield  {journal} {\bibinfo
  {journal} {Physical Review A}\ }\textbf {\bibinfo {volume} {103}},\ \bibinfo
  {pages} {022814} (\bibinfo {year} {2021})}\BibitemShut {NoStop}%
\bibitem [{\citenamefont {Kozlov}\ and\ \citenamefont
  {Labzowsky}(1995)}]{kozlov1995parity}%
  \BibitemOpen
  \bibfield  {author} {\bibinfo {author} {\bibfnamefont {M.~G.}\ \bibnamefont
  {Kozlov}}\ and\ \bibinfo {author} {\bibfnamefont {L.~N.}\ \bibnamefont
  {Labzowsky}},\ }\bibfield  {title} {\bibinfo {title} {Parity violation
  effects in diatomics},\ }\href@noop {} {\bibfield  {journal} {\bibinfo
  {journal} {Journal of Physics B: Atomic, Molecular and Optical Physics}\
  }\textbf {\bibinfo {volume} {28}},\ \bibinfo {pages} {1933} (\bibinfo {year}
  {1995})}\BibitemShut {NoStop}%
\bibitem [{\citenamefont {Kozlov}\ and\ \citenamefont
  {Ezhov}(1994)}]{kozlov1994enhancement}%
  \BibitemOpen
  \bibfield  {author} {\bibinfo {author} {\bibfnamefont {M.}~\bibnamefont
  {Kozlov}}\ and\ \bibinfo {author} {\bibfnamefont {V.}~\bibnamefont {Ezhov}},\
  }\bibfield  {title} {\bibinfo {title} {Enhancement of the electric dipole
  moment of the electron in the {YbF} molecule},\ }\href@noop {} {\bibfield
  {journal} {\bibinfo  {journal} {Physical Review A}\ }\textbf {\bibinfo
  {volume} {49}},\ \bibinfo {pages} {4502} (\bibinfo {year}
  {1994})}\BibitemShut {NoStop}%
\bibitem [{\citenamefont {Haase}\ \emph {et~al.}(2021)\citenamefont {Haase},
  \citenamefont {Doeglas}, \citenamefont {Boeschoten}, \citenamefont {Eliav},
  \citenamefont {Ilia{\v{s}}}, \citenamefont {Aggarwal}, \citenamefont
  {Bethlem}, \citenamefont {Borschevsky}, \citenamefont {Esajas}, \citenamefont
  {Hao}, \citenamefont {Hoekstra}, \citenamefont {Marshall}, \citenamefont
  {Meijknecht}, \citenamefont {Mooij}, \citenamefont {Steinebach},
  \citenamefont {Timmermans}, \citenamefont {Touwen}, \citenamefont {Ubachs},
  \citenamefont {Willmann},\ and\ \citenamefont {and}}]{haase2021systematic}%
  \BibitemOpen
  \bibfield  {author} {\bibinfo {author} {\bibfnamefont {P.~A.~B.}\
  \bibnamefont {Haase}}, \bibinfo {author} {\bibfnamefont {D.~J.}\ \bibnamefont
  {Doeglas}}, \bibinfo {author} {\bibfnamefont {A.}~\bibnamefont {Boeschoten}},
  \bibinfo {author} {\bibfnamefont {E.}~\bibnamefont {Eliav}}, \bibinfo
  {author} {\bibfnamefont {M.}~\bibnamefont {Ilia{\v{s}}}}, \bibinfo {author}
  {\bibfnamefont {P.}~\bibnamefont {Aggarwal}}, \bibinfo {author}
  {\bibfnamefont {H.~L.}\ \bibnamefont {Bethlem}}, \bibinfo {author}
  {\bibfnamefont {A.}~\bibnamefont {Borschevsky}}, \bibinfo {author}
  {\bibfnamefont {K.}~\bibnamefont {Esajas}}, \bibinfo {author} {\bibfnamefont
  {Y.}~\bibnamefont {Hao}}, \bibinfo {author} {\bibfnamefont {S.}~\bibnamefont
  {Hoekstra}}, \bibinfo {author} {\bibfnamefont {V.~R.}\ \bibnamefont
  {Marshall}}, \bibinfo {author} {\bibfnamefont {T.~B.}\ \bibnamefont
  {Meijknecht}}, \bibinfo {author} {\bibfnamefont {M.~C.}\ \bibnamefont
  {Mooij}}, \bibinfo {author} {\bibfnamefont {K.}~\bibnamefont {Steinebach}},
  \bibinfo {author} {\bibfnamefont {R.~G.~E.}\ \bibnamefont {Timmermans}},
  \bibinfo {author} {\bibfnamefont {A.~P.}\ \bibnamefont {Touwen}}, \bibinfo
  {author} {\bibfnamefont {W.}~\bibnamefont {Ubachs}}, \bibinfo {author}
  {\bibfnamefont {L.}~\bibnamefont {Willmann}},\ and\ \bibinfo {author}
  {\bibfnamefont {Y.~Y.}\ \bibnamefont {and}},\ }\bibfield  {title} {\bibinfo
  {title} {Systematic study and uncertainty evaluation of {P,T}-odd molecular
  enhancement factors in {BaF}},\ }\href {https://doi.org/10.1063/5.0047344}
  {\bibfield  {journal} {\bibinfo  {journal} {The Journal of Chemical Physics}\
  }\textbf {\bibinfo {volume} {155}},\ \bibinfo {pages} {034309} (\bibinfo
  {year} {2021})}\BibitemShut {NoStop}%
\bibitem [{\citenamefont {Denis}\ \emph {et~al.}(2019)\citenamefont {Denis},
  \citenamefont {Haase}, \citenamefont {Timmermans}, \citenamefont {Eliav},
  \citenamefont {Hutzler},\ and\ \citenamefont
  {Borschevsky}}]{denis2019enhancement}%
  \BibitemOpen
  \bibfield  {author} {\bibinfo {author} {\bibfnamefont {M.}~\bibnamefont
  {Denis}}, \bibinfo {author} {\bibfnamefont {P.~A.}\ \bibnamefont {Haase}},
  \bibinfo {author} {\bibfnamefont {R.~G.}\ \bibnamefont {Timmermans}},
  \bibinfo {author} {\bibfnamefont {E.}~\bibnamefont {Eliav}}, \bibinfo
  {author} {\bibfnamefont {N.~R.}\ \bibnamefont {Hutzler}},\ and\ \bibinfo
  {author} {\bibfnamefont {A.}~\bibnamefont {Borschevsky}},\ }\bibfield
  {title} {\bibinfo {title} {Enhancement factor for the electric dipole moment
  of the electron in the {BaOH} and {YbOH} molecules},\ }\href@noop {}
  {\bibfield  {journal} {\bibinfo  {journal} {Physical Review A}\ }\textbf
  {\bibinfo {volume} {99}},\ \bibinfo {pages} {042512} (\bibinfo {year}
  {2019})}\BibitemShut {NoStop}%
\bibitem [{\citenamefont {Gaul}\ and\ \citenamefont
  {Berger}(2020)}]{gaul2020ab}%
  \BibitemOpen
  \bibfield  {author} {\bibinfo {author} {\bibfnamefont {K.}~\bibnamefont
  {Gaul}}\ and\ \bibinfo {author} {\bibfnamefont {R.}~\bibnamefont {Berger}},\
  }\bibfield  {title} {\bibinfo {title} {Ab initio study of parity and
  time-reversal violation in laser-coolable triatomic molecules},\ }\href@noop
  {} {\bibfield  {journal} {\bibinfo  {journal} {Physical Review A}\ }\textbf
  {\bibinfo {volume} {101}},\ \bibinfo {pages} {012508} (\bibinfo {year}
  {2020})}\BibitemShut {NoStop}%
\bibitem [{\citenamefont {M{\aa}rtensson-Pendrill}\ and\ \citenamefont
  {{\"O}ster}(1987)}]{maartensson1987calculations}%
  \BibitemOpen
  \bibfield  {author} {\bibinfo {author} {\bibfnamefont {A.-M.}\ \bibnamefont
  {M{\aa}rtensson-Pendrill}}\ and\ \bibinfo {author} {\bibfnamefont
  {P.}~\bibnamefont {{\"O}ster}},\ }\bibfield  {title} {\bibinfo {title}
  {Calculations of atomic electric dipole moments},\ }\href@noop {} {\bibfield
  {journal} {\bibinfo  {journal} {Physica Scripta}\ }\textbf {\bibinfo {volume}
  {36}},\ \bibinfo {pages} {444} (\bibinfo {year} {1987})}\BibitemShut
  {NoStop}%
\bibitem [{\citenamefont {Cohen}\ and\ \citenamefont
  {Roothaan}(1965)}]{cohen1965electric}%
  \BibitemOpen
  \bibfield  {author} {\bibinfo {author} {\bibfnamefont {H.~D.}\ \bibnamefont
  {Cohen}}\ and\ \bibinfo {author} {\bibfnamefont {C.}~\bibnamefont
  {Roothaan}},\ }\bibfield  {title} {\bibinfo {title} {Electric dipole
  polarizability of atoms by the {H}artree-—{F}ock method. {I}. {T}heory for
  closed-shell systems},\ }\href@noop {} {\bibfield  {journal} {\bibinfo
  {journal} {The Journal of Chemical Physics}\ }\textbf {\bibinfo {volume}
  {43}},\ \bibinfo {pages} {S34} (\bibinfo {year} {1965})}\BibitemShut
  {NoStop}%
\bibitem [{\citenamefont {Visscher}\ \emph {et~al.}(1998)\citenamefont
  {Visscher}, \citenamefont {Enevoldsen}, \citenamefont {Saue},\ and\
  \citenamefont {Oddershede}}]{visscher1998molecular}%
  \BibitemOpen
  \bibfield  {author} {\bibinfo {author} {\bibfnamefont {L.}~\bibnamefont
  {Visscher}}, \bibinfo {author} {\bibfnamefont {T.}~\bibnamefont
  {Enevoldsen}}, \bibinfo {author} {\bibfnamefont {T.}~\bibnamefont {Saue}},\
  and\ \bibinfo {author} {\bibfnamefont {J.}~\bibnamefont {Oddershede}},\
  }\bibfield  {title} {\bibinfo {title} {Molecular relativistic calculations of
  the electric field gradients at the nuclei in the hydrogen halides},\
  }\href@noop {} {\bibfield  {journal} {\bibinfo  {journal} {The Journal of
  Chemical Physics}\ }\textbf {\bibinfo {volume} {109}},\ \bibinfo {pages}
  {9677} (\bibinfo {year} {1998})}\BibitemShut {NoStop}%
\bibitem [{\citenamefont {Hao}\ \emph {et~al.}(2018)\citenamefont {Hao},
  \citenamefont {Ilia{\v{s}}}, \citenamefont {Eliav}, \citenamefont
  {Schwerdtfeger}, \citenamefont {Flambaum},\ and\ \citenamefont
  {Borschevsky}}]{hao2018nuclear}%
  \BibitemOpen
  \bibfield  {author} {\bibinfo {author} {\bibfnamefont {Y.}~\bibnamefont
  {Hao}}, \bibinfo {author} {\bibfnamefont {M.}~\bibnamefont {Ilia{\v{s}}}},
  \bibinfo {author} {\bibfnamefont {E.}~\bibnamefont {Eliav}}, \bibinfo
  {author} {\bibfnamefont {P.}~\bibnamefont {Schwerdtfeger}}, \bibinfo {author}
  {\bibfnamefont {V.~V.}\ \bibnamefont {Flambaum}},\ and\ \bibinfo {author}
  {\bibfnamefont {A.}~\bibnamefont {Borschevsky}},\ }\bibfield  {title}
  {\bibinfo {title} {Nuclear anapole moment interaction in {BaF} from
  relativistic coupled-cluster theory},\ }\href@noop {} {\bibfield  {journal}
  {\bibinfo  {journal} {Physical Review A}\ }\textbf {\bibinfo {volume} {98}},\
  \bibinfo {pages} {032510} (\bibinfo {year} {2018})}\BibitemShut {NoStop}%
\bibitem [{\citenamefont {Hao}\ \emph {et~al.}(2020)\citenamefont {Hao},
  \citenamefont {Navr\'atil}, \citenamefont {Norrgard}, \citenamefont
  {Ilia\ifmmode~\check{s}\else \v{s}\fi{}}, \citenamefont {Eliav},
  \citenamefont {Timmermans}, \citenamefont {Flambaum},\ and\ \citenamefont
  {Borschevsky}}]{norrgard2020nuclear}%
  \BibitemOpen
  \bibfield  {author} {\bibinfo {author} {\bibfnamefont {Y.}~\bibnamefont
  {Hao}}, \bibinfo {author} {\bibfnamefont {P.}~\bibnamefont {Navr\'atil}},
  \bibinfo {author} {\bibfnamefont {E.~B.}\ \bibnamefont {Norrgard}}, \bibinfo
  {author} {\bibfnamefont {M.}~\bibnamefont {Ilia\ifmmode~\check{s}\else
  \v{s}\fi{}}}, \bibinfo {author} {\bibfnamefont {E.}~\bibnamefont {Eliav}},
  \bibinfo {author} {\bibfnamefont {R.~G.~E.}\ \bibnamefont {Timmermans}},
  \bibinfo {author} {\bibfnamefont {V.~V.}\ \bibnamefont {Flambaum}},\ and\
  \bibinfo {author} {\bibfnamefont {A.}~\bibnamefont {Borschevsky}},\
  }\bibfield  {title} {\bibinfo {title} {Nuclear spin-dependent
  parity-violating effects in light polyatomic molecules},\ }\href
  {https://doi.org/10.1103/PhysRevA.102.052828} {\bibfield  {journal} {\bibinfo
   {journal} {Phys. Rev. A}\ }\textbf {\bibinfo {volume} {102}},\ \bibinfo
  {pages} {052828} (\bibinfo {year} {2020})}\BibitemShut {NoStop}%
\bibitem [{\citenamefont {Haase}\ \emph {et~al.}(2020)\citenamefont {Haase},
  \citenamefont {Eliav}, \citenamefont {Ilia{\v{s}}},\ and\ \citenamefont
  {Borschevsky}}]{haase2020hyperfine}%
  \BibitemOpen
  \bibfield  {author} {\bibinfo {author} {\bibfnamefont {P.~A.~B.}\
  \bibnamefont {Haase}}, \bibinfo {author} {\bibfnamefont {E.}~\bibnamefont
  {Eliav}}, \bibinfo {author} {\bibfnamefont {M.}~\bibnamefont {Ilia{\v{s}}}},\
  and\ \bibinfo {author} {\bibfnamefont {A.}~\bibnamefont {Borschevsky}},\
  }\bibfield  {title} {\bibinfo {title} {Hyperfine structure constants on the
  relativistic coupled cluster level with associated uncertainties},\ }\href
  {https://doi.org/10.1021/acs.jpca.0c00877} {\bibfield  {journal} {\bibinfo
  {journal} {The Journal of Physical Chemistry A}\ }\textbf {\bibinfo {volume}
  {124}},\ \bibinfo {pages} {3157} (\bibinfo {year} {2020})}\BibitemShut
  {NoStop}%
\bibitem [{\citenamefont {Denis}\ \emph {et~al.}(2022)\citenamefont {Denis},
  \citenamefont {Haase}, \citenamefont {Mooij}, \citenamefont {Chamorro},
  \citenamefont {Aggarwal}, \citenamefont {Bethlem}, \citenamefont
  {Boeschoten}, \citenamefont {Borschevsky}, \citenamefont {Esajas},
  \citenamefont {Hao} \emph {et~al.}}]{denis2022benchmarking}%
  \BibitemOpen
  \bibfield  {author} {\bibinfo {author} {\bibfnamefont {M.}~\bibnamefont
  {Denis}}, \bibinfo {author} {\bibfnamefont {P.~A.}\ \bibnamefont {Haase}},
  \bibinfo {author} {\bibfnamefont {M.~C.}\ \bibnamefont {Mooij}}, \bibinfo
  {author} {\bibfnamefont {Y.}~\bibnamefont {Chamorro}}, \bibinfo {author}
  {\bibfnamefont {P.}~\bibnamefont {Aggarwal}}, \bibinfo {author}
  {\bibfnamefont {H.~L.}\ \bibnamefont {Bethlem}}, \bibinfo {author}
  {\bibfnamefont {A.}~\bibnamefont {Boeschoten}}, \bibinfo {author}
  {\bibfnamefont {A.}~\bibnamefont {Borschevsky}}, \bibinfo {author}
  {\bibfnamefont {K.}~\bibnamefont {Esajas}}, \bibinfo {author} {\bibfnamefont
  {Y.}~\bibnamefont {Hao}}, \emph {et~al.},\ }\bibfield  {title} {\bibinfo
  {title} {Benchmarking of the {F}ock space coupled cluster method and
  uncertainty estimation: Magnetic hyperfine interaction in the excited state
  of {BaF}},\ }\href@noop {} {\bibfield  {journal} {\bibinfo  {journal}
  {Physical Review A}\ }\textbf {\bibinfo {volume} {105}},\ \bibinfo {pages}
  {052811} (\bibinfo {year} {2022})}\BibitemShut {NoStop}%
\bibitem [{\citenamefont {Chupp}\ \emph {et~al.}(2019)\citenamefont {Chupp},
  \citenamefont {Fierlinger}, \citenamefont {Ramsey-Musolf},\ and\
  \citenamefont {Singh}}]{chupp2019electric}%
  \BibitemOpen
  \bibfield  {author} {\bibinfo {author} {\bibfnamefont {T.}~\bibnamefont
  {Chupp}}, \bibinfo {author} {\bibfnamefont {P.}~\bibnamefont {Fierlinger}},
  \bibinfo {author} {\bibfnamefont {M.}~\bibnamefont {Ramsey-Musolf}},\ and\
  \bibinfo {author} {\bibfnamefont {J.}~\bibnamefont {Singh}},\ }\bibfield
  {title} {\bibinfo {title} {Electric dipole moments of atoms, molecules,
  nuclei, and particles},\ }\href@noop {} {\bibfield  {journal} {\bibinfo
  {journal} {Reviews of Modern Physics}\ }\textbf {\bibinfo {volume} {91}},\
  \bibinfo {pages} {015001} (\bibinfo {year} {2019})}\BibitemShut {NoStop}%
\bibitem [{DIR()}]{DIRAC19}%
  \BibitemOpen
  \href@noop {} {}\bibinfo {note} {{DIRAC}, a relativistic ab initio electronic
  structure program, Release {DIRAC19} (2019), written by A.~S.~P.~Gomes,
  T.~Saue, L.~Visscher, H.~J.~{\relax Aa}.~Jensen, and R.~Bast, with
  contributions from I.~A.~Aucar, V.~Bakken, K.~G.~Dyall, S.~Dubillard,
  U.~Ekstr{\"o}m, E.~Eliav, T.~Enevoldsen, E.~Fa{\ss}hauer, T.~Fleig,
  O.~Fossgaard, L.~Halbert, E.~D.~Hedeg{\aa}rd, B.~Heimlich--Paris,
  T.~Helgaker, J.~Henriksson, M.~Ilia{\v{s}}, Ch.~R.~Jacob, S.~Knecht,
  S.~Komorovsk{\'y}, O.~Kullie, J.~K.~L{\ae}rdahl, C.~V.~Larsen, Y.~S.~Lee,
  H.~S.~Nataraj, M.~K.~Nayak, P.~Norman, G.~Olejniczak, J.~Olsen,
  J.~M.~H.~Olsen, Y.~C.~Park, J.~K.~Pedersen, M.~Pernpointner, R.~di~Remigio,
  K.~Ruud, P.~Sa{\l}ek, B.~Schimmelpfennig, B.~Senjean, A.~Shee, J.~Sikkema,
  A.~J.~Thorvaldsen, J.~Thyssen, J.~van~Stralen, M.~L.~Vidal, S.~Villaume,
  O.~Visser, T.~Winther, and S.~Yamamoto (available at
  \url{https://doi.org/10.5281/zenodo.3572669}, see also
  \url{http://www.diracprogram.org})}\BibitemShut {NoStop}%
\bibitem [{\citenamefont {Saue}\ \emph {et~al.}(2020)\citenamefont {Saue},
  \citenamefont {Bast}, \citenamefont {Gomes}, \citenamefont {Jensen},
  \citenamefont {Visscher}, \citenamefont {Aucar}, \citenamefont {Di~Remigio},
  \citenamefont {Dyall}, \citenamefont {Eliav}, \citenamefont {Fasshauer} \emph
  {et~al.}}]{saue2020dirac}%
  \BibitemOpen
  \bibfield  {author} {\bibinfo {author} {\bibfnamefont {T.}~\bibnamefont
  {Saue}}, \bibinfo {author} {\bibfnamefont {R.}~\bibnamefont {Bast}}, \bibinfo
  {author} {\bibfnamefont {A.~S.~P.}\ \bibnamefont {Gomes}}, \bibinfo {author}
  {\bibfnamefont {H.~J.~A.}\ \bibnamefont {Jensen}}, \bibinfo {author}
  {\bibfnamefont {L.}~\bibnamefont {Visscher}}, \bibinfo {author}
  {\bibfnamefont {I.~A.}\ \bibnamefont {Aucar}}, \bibinfo {author}
  {\bibfnamefont {R.}~\bibnamefont {Di~Remigio}}, \bibinfo {author}
  {\bibfnamefont {K.~G.}\ \bibnamefont {Dyall}}, \bibinfo {author}
  {\bibfnamefont {E.}~\bibnamefont {Eliav}}, \bibinfo {author} {\bibfnamefont
  {E.}~\bibnamefont {Fasshauer}}, \emph {et~al.},\ }\bibfield  {title}
  {\bibinfo {title} {The {DIRAC} code for relativistic molecular
  calculations},\ }\href@noop {} {\bibfield  {journal} {\bibinfo  {journal}
  {The Journal of Chemical Physics}\ }\textbf {\bibinfo {volume} {152}},\
  \bibinfo {pages} {204104} (\bibinfo {year} {2020})}\BibitemShut {NoStop}%
\bibitem [{\citenamefont {Stanton}\ \emph {et~al.}()\citenamefont {Stanton},
  \citenamefont {Gauss}, \citenamefont {Cheng}, \citenamefont {Harding},
  \citenamefont {Matthews},\ and\ \citenamefont {Szalay}}]{cfour}%
  \BibitemOpen
  \bibfield  {author} {\bibinfo {author} {\bibfnamefont {J.~F.}\ \bibnamefont
  {Stanton}}, \bibinfo {author} {\bibfnamefont {J.}~\bibnamefont {Gauss}},
  \bibinfo {author} {\bibfnamefont {L.}~\bibnamefont {Cheng}}, \bibinfo
  {author} {\bibfnamefont {M.~E.}\ \bibnamefont {Harding}}, \bibinfo {author}
  {\bibfnamefont {D.~A.}\ \bibnamefont {Matthews}},\ and\ \bibinfo {author}
  {\bibfnamefont {P.~G.}\ \bibnamefont {Szalay}},\ }\href@noop {} {\bibinfo
  {title} {{CFOUR, Coupled-Cluster techniques for Computational Chemistry, a
  quantum-chemical program package}}},\ \bibinfo {note} {{W}ith contributions
  from {A}.{A}. {A}uer, {R}.{J}. {B}artlett, {U}. {B}enedikt, {C}. {B}erger,
  {D}.{E}. {B}ernholdt, {S.} {B}laschke, {Y}. {J}. {B}omble, {S.} {B}urger,
  {O}. {C}hristiansen, {D.} Datta, {F}. Engel, {R}. Faber, {J.} {G}reiner, {M}.
  {H}eckert, {O}. {H}eun, {M}. Hilgenberg, {C}. {H}uber, {T}.-{C}. {J}agau,
  {D}. {J}onsson, {J}. {J}us{\'e}lius, {T}. Kirsch, {K}. {K}lein, {G}.{M.}
  Kopper{W}.{J}. {L}auderdale, {F}. {L}ipparini, {T}. {M}etzroth, {L}.{A}.
  {M}{\"u}ck, {D}.{P}. {O}'{N}eill, {T.} {N}ottoli, {D}.{R}. {P}rice, {E}.
  {P}rochnow, {C}. {P}uzzarini, {K}. {R}uud, {F}. {S}chiffmann, {W}.
  {S}chwalbach, {C}. {S}immons, {S}. {S}topkowicz, {A}. {T}ajti, {J}.
  {V}{\'a}zquez, {F}. {W}ang, {J}.{D}. {W}atts and the integral packages
  {MOLECULE} ({J}. {A}lml{\"o}f and {P}.{R}. {T}aylor), {PROPS} ({P}.{R}.
  {T}aylor), {ABACUS} ({T}. {H}elgaker, {H}.{J}. {A}a. {J}ensen, {P}.
  {J}{\o}rgensen, and {J}. {O}lsen), and {ECP} routines by {A}. {V}. {M}itin
  and {C}. van {W}{\"u}llen. {F}or the current version, see
  http://www.cfour.de.}\BibitemShut {Stop}%
\bibitem [{\citenamefont {Matthews}\ \emph {et~al.}(2020)\citenamefont
  {Matthews}, \citenamefont {Cheng}, \citenamefont {Harding}, \citenamefont
  {Lipparini}, \citenamefont {Stopkowicz}, \citenamefont {Jagau}, \citenamefont
  {Szalay}, \citenamefont {Gauss},\ and\ \citenamefont
  {Stanton}}]{matthews2020coupled}%
  \BibitemOpen
  \bibfield  {author} {\bibinfo {author} {\bibfnamefont {D.~A.}\ \bibnamefont
  {Matthews}}, \bibinfo {author} {\bibfnamefont {L.}~\bibnamefont {Cheng}},
  \bibinfo {author} {\bibfnamefont {M.~E.}\ \bibnamefont {Harding}}, \bibinfo
  {author} {\bibfnamefont {F.}~\bibnamefont {Lipparini}}, \bibinfo {author}
  {\bibfnamefont {S.}~\bibnamefont {Stopkowicz}}, \bibinfo {author}
  {\bibfnamefont {T.-C.}\ \bibnamefont {Jagau}}, \bibinfo {author}
  {\bibfnamefont {P.~G.}\ \bibnamefont {Szalay}}, \bibinfo {author}
  {\bibfnamefont {J.}~\bibnamefont {Gauss}},\ and\ \bibinfo {author}
  {\bibfnamefont {J.~F.}\ \bibnamefont {Stanton}},\ }\bibfield  {title}
  {\bibinfo {title} {Coupled-cluster techniques for computational chemistry:
  The program package},\ }\href@noop {} {\bibfield  {journal} {\bibinfo
  {journal} {The Journal of Chemical Physics}\ }\textbf {\bibinfo {volume}
  {152}},\ \bibinfo {pages} {214108} (\bibinfo {year} {2020})}\BibitemShut
  {NoStop}%
\bibitem [{\citenamefont {Visscher}\ \emph {et~al.}(1996)\citenamefont
  {Visscher}, \citenamefont {Lee},\ and\ \citenamefont
  {Dyall}}]{visscher1996formulation}%
  \BibitemOpen
  \bibfield  {author} {\bibinfo {author} {\bibfnamefont {L.}~\bibnamefont
  {Visscher}}, \bibinfo {author} {\bibfnamefont {T.~J.}\ \bibnamefont {Lee}},\
  and\ \bibinfo {author} {\bibfnamefont {K.~G.}\ \bibnamefont {Dyall}},\
  }\bibfield  {title} {\bibinfo {title} {Formulation and implementation of a
  relativistic unrestricted coupled-cluster method including noniterative
  connected triples},\ }\href@noop {} {\bibfield  {journal} {\bibinfo
  {journal} {The Journal of Chemical Physics}\ }\textbf {\bibinfo {volume}
  {105}},\ \bibinfo {pages} {8769} (\bibinfo {year} {1996})}\BibitemShut
  {NoStop}%
\bibitem [{\citenamefont {Visscher}\ \emph {et~al.}(2001)\citenamefont
  {Visscher}, \citenamefont {Eliav},\ and\ \citenamefont
  {Kaldor}}]{visscher2001formulation}%
  \BibitemOpen
  \bibfield  {author} {\bibinfo {author} {\bibfnamefont {L.}~\bibnamefont
  {Visscher}}, \bibinfo {author} {\bibfnamefont {E.}~\bibnamefont {Eliav}},\
  and\ \bibinfo {author} {\bibfnamefont {U.}~\bibnamefont {Kaldor}},\
  }\bibfield  {title} {\bibinfo {title} {Formulation and implementation of the
  relativistic fock-space coupled cluster method for molecules},\ }\href@noop
  {} {\bibfield  {journal} {\bibinfo  {journal} {The Journal of Chemical
  Physics}\ }\textbf {\bibinfo {volume} {115}},\ \bibinfo {pages} {9720}
  (\bibinfo {year} {2001})}\BibitemShut {NoStop}%
\bibitem [{\citenamefont {Dyall}(2009)}]{dyall2009relativistic}%
  \BibitemOpen
  \bibfield  {author} {\bibinfo {author} {\bibfnamefont {K.~G.}\ \bibnamefont
  {Dyall}},\ }\bibfield  {title} {\bibinfo {title} {Relativistic double-zeta,
  triple-zeta, and quadruple-zeta basis sets for the 4s, 5s, 6s, and 7s
  elements},\ }\href@noop {} {\bibfield  {journal} {\bibinfo  {journal} {The
  Journal of Physical Chemistry A}\ }\textbf {\bibinfo {volume} {113}},\
  \bibinfo {pages} {12638} (\bibinfo {year} {2009})}\BibitemShut {NoStop}%
\bibitem [{\citenamefont {Gomes}\ \emph {et~al.}(2010)\citenamefont {Gomes},
  \citenamefont {Dyall},\ and\ \citenamefont
  {Visscher}}]{gomes2010relativistic}%
  \BibitemOpen
  \bibfield  {author} {\bibinfo {author} {\bibfnamefont {A.~S.}\ \bibnamefont
  {Gomes}}, \bibinfo {author} {\bibfnamefont {K.~G.}\ \bibnamefont {Dyall}},\
  and\ \bibinfo {author} {\bibfnamefont {L.}~\bibnamefont {Visscher}},\
  }\bibfield  {title} {\bibinfo {title} {Relativistic double-zeta, triple-zeta,
  and quadruple-zeta basis sets for the lanthanides {La}--{Lu}},\ }\href@noop
  {} {\bibfield  {journal} {\bibinfo  {journal} {Theoretical Chemistry
  Accounts}\ }\textbf {\bibinfo {volume} {127}},\ \bibinfo {pages} {369}
  (\bibinfo {year} {2010})}\BibitemShut {NoStop}%
\bibitem [{\citenamefont {Dyall}(2016)}]{dyall2016relativistic}%
  \BibitemOpen
  \bibfield  {author} {\bibinfo {author} {\bibfnamefont {K.~G.}\ \bibnamefont
  {Dyall}},\ }\bibfield  {title} {\bibinfo {title} {Relativistic double-zeta,
  triple-zeta, and quadruple-zeta basis sets for the light elements
  {H}--{Ar}},\ }\href@noop {} {\bibfield  {journal} {\bibinfo  {journal}
  {Theoretical Chemistry Accounts}\ }\textbf {\bibinfo {volume} {135}},\
  \bibinfo {pages} {128} (\bibinfo {year} {2016})}\BibitemShut {NoStop}%
\bibitem [{\citenamefont {Roos}\ \emph
  {et~al.}(2004{\natexlab{a}})\citenamefont {Roos}, \citenamefont {Veryazov},\
  and\ \citenamefont {Widmark}}]{roos2004relativistic}%
  \BibitemOpen
  \bibfield  {author} {\bibinfo {author} {\bibfnamefont {B.~O.}\ \bibnamefont
  {Roos}}, \bibinfo {author} {\bibfnamefont {V.}~\bibnamefont {Veryazov}},\
  and\ \bibinfo {author} {\bibfnamefont {P.-O.}\ \bibnamefont {Widmark}},\
  }\bibfield  {title} {\bibinfo {title} {Relativistic atomic natural orbital
  type basis sets for the alkaline and alkaline-earth atoms applied to the
  ground-state potentials for the corresponding dimers},\ }\href@noop {}
  {\bibfield  {journal} {\bibinfo  {journal} {Theoretical Chemistry Accounts}\
  }\textbf {\bibinfo {volume} {111}},\ \bibinfo {pages} {345} (\bibinfo {year}
  {2004}{\natexlab{a}})}\BibitemShut {NoStop}%
\bibitem [{\citenamefont {Roos}\ \emph
  {et~al.}(2004{\natexlab{b}})\citenamefont {Roos}, \citenamefont {Lindh},
  \citenamefont {Malmqvist}, \citenamefont {Veryazov},\ and\ \citenamefont
  {Widmark}}]{roos2004main}%
  \BibitemOpen
  \bibfield  {author} {\bibinfo {author} {\bibfnamefont {B.~O.}\ \bibnamefont
  {Roos}}, \bibinfo {author} {\bibfnamefont {R.}~\bibnamefont {Lindh}},
  \bibinfo {author} {\bibfnamefont {P.-{\AA}.}\ \bibnamefont {Malmqvist}},
  \bibinfo {author} {\bibfnamefont {V.}~\bibnamefont {Veryazov}},\ and\
  \bibinfo {author} {\bibfnamefont {P.-O.}\ \bibnamefont {Widmark}},\
  }\bibfield  {title} {\bibinfo {title} {Main group atoms and dimers studied
  with a new relativistic {ANO} basis set},\ }\href@noop {} {\bibfield
  {journal} {\bibinfo  {journal} {The Journal of Physical Chemistry A}\
  }\textbf {\bibinfo {volume} {108}},\ \bibinfo {pages} {2851} (\bibinfo {year}
  {2004}{\natexlab{b}})}\BibitemShut {NoStop}%
\bibitem [{\citenamefont {Roos}\ \emph {et~al.}(2008)\citenamefont {Roos},
  \citenamefont {Lindh}, \citenamefont {Malmqvist}, \citenamefont {Veryazov},
  \citenamefont {Widmark},\ and\ \citenamefont {Borin}}]{roos2008new}%
  \BibitemOpen
  \bibfield  {author} {\bibinfo {author} {\bibfnamefont {B.~O.}\ \bibnamefont
  {Roos}}, \bibinfo {author} {\bibfnamefont {R.}~\bibnamefont {Lindh}},
  \bibinfo {author} {\bibfnamefont {P.-{\AA}.}\ \bibnamefont {Malmqvist}},
  \bibinfo {author} {\bibfnamefont {V.}~\bibnamefont {Veryazov}}, \bibinfo
  {author} {\bibfnamefont {P.-O.}\ \bibnamefont {Widmark}},\ and\ \bibinfo
  {author} {\bibfnamefont {A.~C.}\ \bibnamefont {Borin}},\ }\bibfield  {title}
  {\bibinfo {title} {New relativistic atomic natural orbital basis sets for
  lanthanide atoms with applications to the ce diatom and {LuF$_3$}},\
  }\href@noop {} {\bibfield  {journal} {\bibinfo  {journal} {The Journal of
  Physical Chemistry A}\ }\textbf {\bibinfo {volume} {112}},\ \bibinfo {pages}
  {11431} (\bibinfo {year} {2008})}\BibitemShut {NoStop}%
\bibitem [{\citenamefont {Widmark}\ \emph {et~al.}(1990)\citenamefont
  {Widmark}, \citenamefont {Malmqvist},\ and\ \citenamefont
  {Roos}}]{widmark1990density}%
  \BibitemOpen
  \bibfield  {author} {\bibinfo {author} {\bibfnamefont {P.-O.}\ \bibnamefont
  {Widmark}}, \bibinfo {author} {\bibfnamefont {P.-{\AA}.}\ \bibnamefont
  {Malmqvist}},\ and\ \bibinfo {author} {\bibfnamefont {B.~O.}\ \bibnamefont
  {Roos}},\ }\bibfield  {title} {\bibinfo {title} {Density matrix averaged
  atomic natural orbital ({ANO}) basis sets for correlated molecular wave
  functions},\ }\href@noop {} {\bibfield  {journal} {\bibinfo  {journal}
  {Theoretica chimica acta}\ }\textbf {\bibinfo {volume} {77}},\ \bibinfo
  {pages} {291} (\bibinfo {year} {1990})}\BibitemShut {NoStop}%
\bibitem [{\citenamefont {Xin}\ \emph {et~al.}(1998{\natexlab{a}})\citenamefont
  {Xin}, \citenamefont {Robinson}, \citenamefont {Apponi},\ and\ \citenamefont
  {Ziurys}}]{xin1998high}%
  \BibitemOpen
  \bibfield  {author} {\bibinfo {author} {\bibfnamefont {J.}~\bibnamefont
  {Xin}}, \bibinfo {author} {\bibfnamefont {J.}~\bibnamefont {Robinson}},
  \bibinfo {author} {\bibfnamefont {A.}~\bibnamefont {Apponi}},\ and\ \bibinfo
  {author} {\bibfnamefont {L.~M.}\ \bibnamefont {Ziurys}},\ }\bibfield  {title}
  {\bibinfo {title} {High resolution spectroscopy of {BaCH}$_3$
  ({$\tilde{\mathrm{X}}^2$}{A}$_1$): Fine and hyperfine structure analysis},\
  }\href@noop {} {\bibfield  {journal} {\bibinfo  {journal} {The Journal of
  Chemical Physics}\ }\textbf {\bibinfo {volume} {108}},\ \bibinfo {pages}
  {2703} (\bibinfo {year} {1998}{\natexlab{a}})}\BibitemShut {NoStop}%
\bibitem [{\citenamefont {Schiff}(1963)}]{schiff1963measurability}%
  \BibitemOpen
  \bibfield  {author} {\bibinfo {author} {\bibfnamefont {L.}~\bibnamefont
  {Schiff}},\ }\bibfield  {title} {\bibinfo {title} {Measurability of nuclear
  electric dipole moments},\ }\href@noop {} {\bibfield  {journal} {\bibinfo
  {journal} {Physical Review}\ }\textbf {\bibinfo {volume} {132}},\ \bibinfo
  {pages} {2194} (\bibinfo {year} {1963})}\BibitemShut {NoStop}%
\bibitem [{\citenamefont {Pa{\v{s}}teka}\ \emph {et~al.}(2016)\citenamefont
  {Pa{\v{s}}teka}, \citenamefont {Mawhorter},\ and\ \citenamefont
  {Schwerdtfeger}}]{pavsteka2016relativistic}%
  \BibitemOpen
  \bibfield  {author} {\bibinfo {author} {\bibfnamefont {L.}~\bibnamefont
  {Pa{\v{s}}teka}}, \bibinfo {author} {\bibfnamefont {R.}~\bibnamefont
  {Mawhorter}},\ and\ \bibinfo {author} {\bibfnamefont {P.}~\bibnamefont
  {Schwerdtfeger}},\ }\bibfield  {title} {\bibinfo {title} {Relativistic
  coupled-cluster calculations of the $^{173}{Yb}$ nuclear quadrupole coupling
  constant for the {YbF} molecule},\ }\href@noop {} {\bibfield  {journal}
  {\bibinfo  {journal} {Molecular Physics}\ }\textbf {\bibinfo {volume}
  {114}},\ \bibinfo {pages} {1110} (\bibinfo {year} {2016})}\BibitemShut
  {NoStop}%
\bibitem [{\citenamefont {Lee}\ and\ \citenamefont
  {Taylor}(1989)}]{lee1989diagnostic}%
  \BibitemOpen
  \bibfield  {author} {\bibinfo {author} {\bibfnamefont {T.~J.}\ \bibnamefont
  {Lee}}\ and\ \bibinfo {author} {\bibfnamefont {P.~R.}\ \bibnamefont
  {Taylor}},\ }\bibfield  {title} {\bibinfo {title} {A diagnostic for
  determining the quality of single-reference electron correlation methods},\
  }\href@noop {} {\bibfield  {journal} {\bibinfo  {journal} {International
  Journal of Quantum Chemistry}\ }\textbf {\bibinfo {volume} {36}},\ \bibinfo
  {pages} {199} (\bibinfo {year} {1989})}\BibitemShut {NoStop}%
\bibitem [{\citenamefont {Dunning~Jr}(1989)}]{dunning1989gaussian}%
  \BibitemOpen
  \bibfield  {author} {\bibinfo {author} {\bibfnamefont {T.~H.}\ \bibnamefont
  {Dunning~Jr}},\ }\bibfield  {title} {\bibinfo {title} {Gaussian basis sets
  for use in correlated molecular calculations. {I}. {T}he atoms boron through
  neon and hydrogen},\ }\href@noop {} {\bibfield  {journal} {\bibinfo
  {journal} {The Journal of Chemical Physics}\ }\textbf {\bibinfo {volume}
  {90}},\ \bibinfo {pages} {1007} (\bibinfo {year} {1989})}\BibitemShut
  {NoStop}%
\bibitem [{\citenamefont {Feller}(1992)}]{feller1992application}%
  \BibitemOpen
  \bibfield  {author} {\bibinfo {author} {\bibfnamefont {D.}~\bibnamefont
  {Feller}},\ }\bibfield  {title} {\bibinfo {title} {Application of systematic
  sequences of wave functions to the water dimer},\ }\href@noop {} {\bibfield
  {journal} {\bibinfo  {journal} {The Journal of Chemical Physics}\ }\textbf
  {\bibinfo {volume} {96}},\ \bibinfo {pages} {6104} (\bibinfo {year}
  {1992})}\BibitemShut {NoStop}%
\bibitem [{\citenamefont {Helgaker}\ \emph {et~al.}(1997)\citenamefont
  {Helgaker}, \citenamefont {Klopper}, \citenamefont {Koch},\ and\
  \citenamefont {Noga}}]{helgaker1997basis}%
  \BibitemOpen
  \bibfield  {author} {\bibinfo {author} {\bibfnamefont {T.}~\bibnamefont
  {Helgaker}}, \bibinfo {author} {\bibfnamefont {W.}~\bibnamefont {Klopper}},
  \bibinfo {author} {\bibfnamefont {H.}~\bibnamefont {Koch}},\ and\ \bibinfo
  {author} {\bibfnamefont {J.}~\bibnamefont {Noga}},\ }\bibfield  {title}
  {\bibinfo {title} {Basis-set convergence of correlated calculations on
  water},\ }\href@noop {} {\bibfield  {journal} {\bibinfo  {journal} {The
  Journal of Chemical Physics}\ }\textbf {\bibinfo {volume} {106}},\ \bibinfo
  {pages} {9639} (\bibinfo {year} {1997})}\BibitemShut {NoStop}%
\bibitem [{\citenamefont {Martin}(1996)}]{martin1996ab}%
  \BibitemOpen
  \bibfield  {author} {\bibinfo {author} {\bibfnamefont {J.~M.}\ \bibnamefont
  {Martin}},\ }\bibfield  {title} {\bibinfo {title} {Ab initio total
  atomization energies of small molecules—towards the basis set limit},\
  }\href@noop {} {\bibfield  {journal} {\bibinfo  {journal} {Chemical Physics
  letters}\ }\textbf {\bibinfo {volume} {259}},\ \bibinfo {pages} {669}
  (\bibinfo {year} {1996})}\BibitemShut {NoStop}%
\bibitem [{\citenamefont {Lesiuk}\ and\ \citenamefont
  {Jeziorski}(2019)}]{lesiuk2019complete}%
  \BibitemOpen
  \bibfield  {author} {\bibinfo {author} {\bibfnamefont {M.}~\bibnamefont
  {Lesiuk}}\ and\ \bibinfo {author} {\bibfnamefont {B.}~\bibnamefont
  {Jeziorski}},\ }\bibfield  {title} {\bibinfo {title} {Complete basis set
  extrapolation of electronic correlation energies using the {R}iemann zeta
  function},\ }\href@noop {} {\bibfield  {journal} {\bibinfo  {journal}
  {Journal of Chemical Theory and Computation}\ }\textbf {\bibinfo {volume}
  {15}},\ \bibinfo {pages} {5398} (\bibinfo {year} {2019})}\BibitemShut
  {NoStop}%
\bibitem [{\citenamefont {Guo}\ \emph {et~al.}(2021)\citenamefont {Guo},
  \citenamefont {Pa{\v{s}}teka}, \citenamefont {Eliav},\ and\ \citenamefont
  {Borschevsky}}]{guo2021ionization}%
  \BibitemOpen
  \bibfield  {author} {\bibinfo {author} {\bibfnamefont {Y.}~\bibnamefont
  {Guo}}, \bibinfo {author} {\bibfnamefont {L.~F.}\ \bibnamefont
  {Pa{\v{s}}teka}}, \bibinfo {author} {\bibfnamefont {E.}~\bibnamefont
  {Eliav}},\ and\ \bibinfo {author} {\bibfnamefont {A.}~\bibnamefont
  {Borschevsky}},\ }\bibfield  {title} {\bibinfo {title} {Ionization potentials
  and electron affinity of oganesson with relativistic coupled cluster
  method},\ }in\ \href@noop {} {\emph {\bibinfo {booktitle} {Advances in
  Quantum Chemistry}}},\ Vol.~\bibinfo {volume} {83}\ (\bibinfo  {publisher}
  {Elsevier},\ \bibinfo {year} {2021})\ pp.\ \bibinfo {pages}
  {107--123}\BibitemShut {NoStop}%
\bibitem [{\citenamefont {Lindroth}\ \emph {et~al.}(1989)\citenamefont
  {Lindroth}, \citenamefont {Lynn},\ and\ \citenamefont
  {Sandars}}]{lindroth1989order}%
  \BibitemOpen
  \bibfield  {author} {\bibinfo {author} {\bibfnamefont {E.}~\bibnamefont
  {Lindroth}}, \bibinfo {author} {\bibfnamefont {B.}~\bibnamefont {Lynn}},\
  and\ \bibinfo {author} {\bibfnamefont {P.}~\bibnamefont {Sandars}},\
  }\bibfield  {title} {\bibinfo {title} {Order $\alpha^2$ theory of the atomic
  electric dipole moment due to an electric dipole moment on the electron},\
  }\href@noop {} {\bibfield  {journal} {\bibinfo  {journal} {Journal of Physics
  B: Atomic, Molecular and Optical Physics}\ }\textbf {\bibinfo {volume}
  {22}},\ \bibinfo {pages} {559} (\bibinfo {year} {1989})}\BibitemShut
  {NoStop}%
\bibitem [{\citenamefont {Gaunt}(1929)}]{gaunt1929iv}%
  \BibitemOpen
  \bibfield  {author} {\bibinfo {author} {\bibfnamefont {J.~A.}\ \bibnamefont
  {Gaunt}},\ }\bibfield  {title} {\bibinfo {title} {{IV}. {T}he triplets of
  helium},\ }\href@noop {} {\bibfield  {journal} {\bibinfo  {journal}
  {Philosophical Transactions of the Royal Society of London. Series A,
  Containing Papers of a Mathematical or Physical Character}\ }\textbf
  {\bibinfo {volume} {228}},\ \bibinfo {pages} {151} (\bibinfo {year}
  {1929})}\BibitemShut {NoStop}%
\bibitem [{\citenamefont {Zhang}\ \emph {et~al.}(2021)\citenamefont {Zhang},
  \citenamefont {Zheng},\ and\ \citenamefont {Cheng}}]{zhang2021calculations}%
  \BibitemOpen
  \bibfield  {author} {\bibinfo {author} {\bibfnamefont {C.}~\bibnamefont
  {Zhang}}, \bibinfo {author} {\bibfnamefont {X.}~\bibnamefont {Zheng}},\ and\
  \bibinfo {author} {\bibfnamefont {L.}~\bibnamefont {Cheng}},\ }\bibfield
  {title} {\bibinfo {title} {Calculations of time-reversal-symmetry-violation
  sensitivity parameters based on analytic relativistic coupled-cluster
  gradient theory},\ }\href@noop {} {\bibfield  {journal} {\bibinfo  {journal}
  {Physical Review A}\ }\textbf {\bibinfo {volume} {104}},\ \bibinfo {pages}
  {012814} (\bibinfo {year} {2021})}\BibitemShut {NoStop}%
\bibitem [{\citenamefont {Prasannaa}\ \emph {et~al.}(2019)\citenamefont
  {Prasannaa}, \citenamefont {Shitara}, \citenamefont {Sakurai}, \citenamefont
  {Abe},\ and\ \citenamefont {Das}}]{prasannaa2019enhanced}%
  \BibitemOpen
  \bibfield  {author} {\bibinfo {author} {\bibfnamefont {V.}~\bibnamefont
  {Prasannaa}}, \bibinfo {author} {\bibfnamefont {N.}~\bibnamefont {Shitara}},
  \bibinfo {author} {\bibfnamefont {A.}~\bibnamefont {Sakurai}}, \bibinfo
  {author} {\bibfnamefont {M.}~\bibnamefont {Abe}},\ and\ \bibinfo {author}
  {\bibfnamefont {B.}~\bibnamefont {Das}},\ }\bibfield  {title} {\bibinfo
  {title} {Enhanced sensitivity of the electron electric dipole moment from
  {YbOH}: The role of theory},\ }\href@noop {} {\bibfield  {journal} {\bibinfo
  {journal} {Physical Review A}\ }\textbf {\bibinfo {volume} {99}},\ \bibinfo
  {pages} {062502} (\bibinfo {year} {2019})}\BibitemShut {NoStop}%
\bibitem [{\citenamefont {Bader}\ and\ \citenamefont
  {Nguyen-Dang}(1981)}]{bader1981quantum}%
  \BibitemOpen
  \bibfield  {author} {\bibinfo {author} {\bibfnamefont {R.~F.}\ \bibnamefont
  {Bader}}\ and\ \bibinfo {author} {\bibfnamefont {T.}~\bibnamefont
  {Nguyen-Dang}},\ }\bibfield  {title} {\bibinfo {title} {Quantum theory of
  atoms in molecules--dalton revisited},\ }in\ \href@noop {} {\emph {\bibinfo
  {booktitle} {Advances in Quantum Chemistry}}},\ Vol.~\bibinfo {volume} {14}\
  (\bibinfo  {publisher} {Elsevier},\ \bibinfo {year} {1981})\ pp.\ \bibinfo
  {pages} {63--124}\BibitemShut {NoStop}%
\bibitem [{\citenamefont {Glendening}\ \emph {et~al.}(2013)\citenamefont
  {Glendening}, \citenamefont {Landis},\ and\ \citenamefont
  {Weinhold}}]{glendening2013nbo}%
  \BibitemOpen
  \bibfield  {author} {\bibinfo {author} {\bibfnamefont {E.~D.}\ \bibnamefont
  {Glendening}}, \bibinfo {author} {\bibfnamefont {C.~R.}\ \bibnamefont
  {Landis}},\ and\ \bibinfo {author} {\bibfnamefont {F.}~\bibnamefont
  {Weinhold}},\ }\bibfield  {title} {\bibinfo {title} {Nbo 6.0: Natural bond
  orbital analysis program},\ }\href@noop {} {\bibfield  {journal} {\bibinfo
  {journal} {Journal of computational chemistry}\ }\textbf {\bibinfo {volume}
  {34}},\ \bibinfo {pages} {1429} (\bibinfo {year} {2013})}\BibitemShut
  {NoStop}%
\bibitem [{\citenamefont {Perdew}\ \emph {et~al.}(1996)\citenamefont {Perdew},
  \citenamefont {Burke},\ and\ \citenamefont
  {Ernzerhof}}]{PhysRevLett.77.3865}%
  \BibitemOpen
  \bibfield  {author} {\bibinfo {author} {\bibfnamefont {J.~P.}\ \bibnamefont
  {Perdew}}, \bibinfo {author} {\bibfnamefont {K.}~\bibnamefont {Burke}},\ and\
  \bibinfo {author} {\bibfnamefont {M.}~\bibnamefont {Ernzerhof}},\ }\bibfield
  {title} {\bibinfo {title} {Generalized gradient approximation made simple},\
  }\href {https://doi.org/10.1103/PhysRevLett.77.3865} {\bibfield  {journal}
  {\bibinfo  {journal} {Phys. Rev. Lett.}\ }\textbf {\bibinfo {volume} {77}},\
  \bibinfo {pages} {3865} (\bibinfo {year} {1996})}\BibitemShut {NoStop}%
\bibitem [{\citenamefont {Te~Velde}\ \emph {et~al.}(2001)\citenamefont
  {Te~Velde}, \citenamefont {Bickelhaupt}, \citenamefont {Baerends},
  \citenamefont {Fonseca~Guerra}, \citenamefont {van Gisbergen}, \citenamefont
  {Snijders},\ and\ \citenamefont {Ziegler}}]{te2001chemistry}%
  \BibitemOpen
  \bibfield  {author} {\bibinfo {author} {\bibfnamefont {G.~t.}\ \bibnamefont
  {Te~Velde}}, \bibinfo {author} {\bibfnamefont {F.~M.}\ \bibnamefont
  {Bickelhaupt}}, \bibinfo {author} {\bibfnamefont {E.~J.}\ \bibnamefont
  {Baerends}}, \bibinfo {author} {\bibfnamefont {C.}~\bibnamefont
  {Fonseca~Guerra}}, \bibinfo {author} {\bibfnamefont {S.~J.}\ \bibnamefont
  {van Gisbergen}}, \bibinfo {author} {\bibfnamefont {J.~G.}\ \bibnamefont
  {Snijders}},\ and\ \bibinfo {author} {\bibfnamefont {T.}~\bibnamefont
  {Ziegler}},\ }\bibfield  {title} {\bibinfo {title} {Chemistry with adf},\
  }\href@noop {} {\bibfield  {journal} {\bibinfo  {journal} {Journal of
  Computational Chemistry}\ }\textbf {\bibinfo {volume} {22}},\ \bibinfo
  {pages} {931} (\bibinfo {year} {2001})}\BibitemShut {NoStop}%
\bibitem [{\citenamefont {Espinosa}\ \emph {et~al.}(2002)\citenamefont
  {Espinosa}, \citenamefont {Alkorta}, \citenamefont {Elguero},\ and\
  \citenamefont {Molins}}]{espinosa2002weak}%
  \BibitemOpen
  \bibfield  {author} {\bibinfo {author} {\bibfnamefont {E.}~\bibnamefont
  {Espinosa}}, \bibinfo {author} {\bibfnamefont {I.}~\bibnamefont {Alkorta}},
  \bibinfo {author} {\bibfnamefont {J.}~\bibnamefont {Elguero}},\ and\ \bibinfo
  {author} {\bibfnamefont {E.}~\bibnamefont {Molins}},\ }\bibfield  {title}
  {\bibinfo {title} {From weak to strong interactions: A comprehensive analysis
  of the topological and energetic properties of the electron density
  distribution involving {X--H}$\cdot\cdot\cdot$ {F--Y} systems},\ }\href@noop
  {} {\bibfield  {journal} {\bibinfo  {journal} {The Journal of chemical
  physics}\ }\textbf {\bibinfo {volume} {117}},\ \bibinfo {pages} {5529}
  (\bibinfo {year} {2002})}\BibitemShut {NoStop}%
\bibitem [{\citenamefont {Sunaga}\ \emph {et~al.}(2017)\citenamefont {Sunaga},
  \citenamefont {Abe}, \citenamefont {Hada},\ and\ \citenamefont
  {Das}}]{sunaga2017analysis}%
  \BibitemOpen
  \bibfield  {author} {\bibinfo {author} {\bibfnamefont {A.}~\bibnamefont
  {Sunaga}}, \bibinfo {author} {\bibfnamefont {M.}~\bibnamefont {Abe}},
  \bibinfo {author} {\bibfnamefont {M.}~\bibnamefont {Hada}},\ and\ \bibinfo
  {author} {\bibfnamefont {B.}~\bibnamefont {Das}},\ }\bibfield  {title}
  {\bibinfo {title} {Analysis of large effective electric fields of weakly
  polar molecules for electron electric-dipole-moment searches},\ }\href@noop
  {} {\bibfield  {journal} {\bibinfo  {journal} {Physical Review A}\ }\textbf
  {\bibinfo {volume} {95}},\ \bibinfo {pages} {012502} (\bibinfo {year}
  {2017})}\BibitemShut {NoStop}%
\bibitem [{\citenamefont {Meyer}\ \emph {et~al.}(2006)\citenamefont {Meyer},
  \citenamefont {Bohn},\ and\ \citenamefont {Deskevich}}]{meyer2006candidate}%
  \BibitemOpen
  \bibfield  {author} {\bibinfo {author} {\bibfnamefont {E.~R.}\ \bibnamefont
  {Meyer}}, \bibinfo {author} {\bibfnamefont {J.~L.}\ \bibnamefont {Bohn}},\
  and\ \bibinfo {author} {\bibfnamefont {M.~P.}\ \bibnamefont {Deskevich}},\
  }\bibfield  {title} {\bibinfo {title} {Candidate molecular ions for an
  electron electric dipole moment experiment},\ }\href@noop {} {\bibfield
  {journal} {\bibinfo  {journal} {Physical Review A}\ }\textbf {\bibinfo
  {volume} {73}},\ \bibinfo {pages} {062108} (\bibinfo {year}
  {2006})}\BibitemShut {NoStop}%
\bibitem [{\citenamefont {Hutzler}\ \emph {et~al.}(2012)\citenamefont
  {Hutzler}, \citenamefont {Lu},\ and\ \citenamefont {Doyle}}]{Hutzler2012}%
  \BibitemOpen
  \bibfield  {author} {\bibinfo {author} {\bibfnamefont {N.~R.}\ \bibnamefont
  {Hutzler}}, \bibinfo {author} {\bibfnamefont {H.-I.}\ \bibnamefont {Lu}},\
  and\ \bibinfo {author} {\bibfnamefont {J.~M.}\ \bibnamefont {Doyle}},\
  }\bibfield  {title} {\bibinfo {title} {{The buffer gas beam: an intense,
  cold, and slow source for atoms and molecules.}},\ }\href
  {https://doi.org/10.1021/cr200362u} {\bibfield  {journal} {\bibinfo
  {journal} {Chemical Reviews}\ }\textbf {\bibinfo {volume} {112}},\ \bibinfo
  {pages} {4803} (\bibinfo {year} {2012})}\BibitemShut {NoStop}%
\bibitem [{\citenamefont {Xin}\ \emph {et~al.}(1998{\natexlab{b}})\citenamefont
  {Xin}, \citenamefont {Robinson}, \citenamefont {Apponi},\ and\ \citenamefont
  {Ziurys}}]{Xin1998}%
  \BibitemOpen
  \bibfield  {author} {\bibinfo {author} {\bibfnamefont {J.}~\bibnamefont
  {Xin}}, \bibinfo {author} {\bibfnamefont {J.~S.}\ \bibnamefont {Robinson}},
  \bibinfo {author} {\bibfnamefont {A.~J.}\ \bibnamefont {Apponi}},\ and\
  \bibinfo {author} {\bibfnamefont {L.~M.}\ \bibnamefont {Ziurys}},\ }\bibfield
   {title} {\bibinfo {title} {High resolution spectroscopy of {BaCH$_3$($
  \tilde{X}^2A_1$}): Fine and hyperfine structure analysis},\ }\href
  {https://doi.org/10.1063/1.475662} {\bibfield  {journal} {\bibinfo  {journal}
  {The Journal of Chemical Physics}\ }\textbf {\bibinfo {volume} {108}},\
  \bibinfo {pages} {2703} (\bibinfo {year} {1998}{\natexlab{b}})}\BibitemShut
  {NoStop}%
\bibitem [{\citenamefont {Brazier}\ and\ \citenamefont
  {Bernath}(1987)}]{Brazier1987}%
  \BibitemOpen
  \bibfield  {author} {\bibinfo {author} {\bibfnamefont {C.~R.}\ \bibnamefont
  {Brazier}}\ and\ \bibinfo {author} {\bibfnamefont {P.~F.}\ \bibnamefont
  {Bernath}},\ }\bibfield  {title} {\bibinfo {title} {Observation of gas phase
  organometallic free radicals: Monomethyl derivatives of calcium and
  strontium},\ }\href {https://doi.org/10.1063/1.452476} {\bibfield  {journal}
  {\bibinfo  {journal} {The Journal of Chemical Physics}\ }\textbf {\bibinfo
  {volume} {86}},\ \bibinfo {pages} {5918} (\bibinfo {year}
  {1987})}\BibitemShut {NoStop}%
\bibitem [{\citenamefont {Jadbabaie}\ \emph {et~al.}(2020)\citenamefont
  {Jadbabaie}, \citenamefont {Pilgram}, \citenamefont {K\l~os}, \citenamefont
  {Kotochigova},\ and\ \citenamefont {Hutzler}}]{Jadbabaie2020}%
  \BibitemOpen
  \bibfield  {author} {\bibinfo {author} {\bibfnamefont {A.}~\bibnamefont
  {Jadbabaie}}, \bibinfo {author} {\bibfnamefont {N.~H.}\ \bibnamefont
  {Pilgram}}, \bibinfo {author} {\bibfnamefont {J.}~\bibnamefont {K\l~os}},
  \bibinfo {author} {\bibfnamefont {S.}~\bibnamefont {Kotochigova}},\ and\
  \bibinfo {author} {\bibfnamefont {N.~R.}\ \bibnamefont {Hutzler}},\
  }\bibfield  {title} {\bibinfo {title} {{Enhanced molecular yield from a
  cryogenic buffer gas beam source via excited state chemistry}},\ }\href
  {https://doi.org/10.1088/1367-2630/ab6eae} {\bibfield  {journal} {\bibinfo
  {journal} {New Journal of Physics}\ }\textbf {\bibinfo {volume} {22}},\
  \bibinfo {pages} {022002} (\bibinfo {year} {2020})}\BibitemShut {NoStop}%
\bibitem [{\citenamefont {Yu}\ and\ \citenamefont {Hutzler}(2021)}]{Yu2021}%
  \BibitemOpen
  \bibfield  {author} {\bibinfo {author} {\bibfnamefont {P.}~\bibnamefont
  {Yu}}\ and\ \bibinfo {author} {\bibfnamefont {N.~R.}\ \bibnamefont
  {Hutzler}},\ }\bibfield  {title} {\bibinfo {title} {Probing fundamental
  symmetries of deformed nuclei in symmetric top molecules},\ }\bibfield
  {journal} {\bibinfo  {journal} {Physical Review Letters}\ }\textbf {\bibinfo
  {volume} {126}},\ \href {https://doi.org/10.1103/physrevlett.126.023003}
  {10.1103/physrevlett.126.023003} (\bibinfo {year} {2021})\BibitemShut
  {NoStop}%
\bibitem [{\citenamefont {ichi C.~Namiki}\ and\ \citenamefont
  {Steimle}(1999)}]{Namiki1999}%
  \BibitemOpen
  \bibfield  {author} {\bibinfo {author} {\bibfnamefont {K.}~\bibnamefont {ichi
  C.~Namiki}}\ and\ \bibinfo {author} {\bibfnamefont {T.~C.}\ \bibnamefont
  {Steimle}},\ }\bibfield  {title} {\bibinfo {title} {Pure rotational spectrum
  of {CaCH$_3$($\tilde{X}^2A_1$)} using the pump/probe microwave-optical double
  resonance ({PPMODR}) technique},\ }\href {https://doi.org/10.1063/1.479071}
  {\bibfield  {journal} {\bibinfo  {journal} {The Journal of Chemical Physics}\
  }\textbf {\bibinfo {volume} {110}},\ \bibinfo {pages} {11309} (\bibinfo
  {year} {1999})}\BibitemShut {NoStop}%
\bibitem [{\citenamefont {Sheridan}\ \emph {et~al.}(2005)\citenamefont
  {Sheridan}, \citenamefont {Dick}, \citenamefont {Wang},\ and\ \citenamefont
  {Bernath}}]{Sheridan2005}%
  \BibitemOpen
  \bibfield  {author} {\bibinfo {author} {\bibfnamefont {P.~M.}\ \bibnamefont
  {Sheridan}}, \bibinfo {author} {\bibfnamefont {M.~J.}\ \bibnamefont {Dick}},
  \bibinfo {author} {\bibfnamefont {J.-G.}\ \bibnamefont {Wang}},\ and\
  \bibinfo {author} {\bibfnamefont {P.~F.}\ \bibnamefont {Bernath}},\
  }\bibfield  {title} {\bibinfo {title} {High-resolution spectroscopic
  investigation of the {$\tilde{B}^2A_1-\tilde{X}^2A_1$} transitions of
  {CaCH$_3$} and {SrCH$_3$}},\ }\href {https://doi.org/10.1021/jp054228q}
  {\bibfield  {journal} {\bibinfo  {journal} {The Journal of Physical Chemistry
  A}\ }\textbf {\bibinfo {volume} {109}},\ \bibinfo {pages} {10547} (\bibinfo
  {year} {2005})}\BibitemShut {NoStop}%
\bibitem [{\citenamefont {ichi C.~Namiki}\ \emph {et~al.}(1998)\citenamefont
  {ichi C.~Namiki}, \citenamefont {Robinson},\ and\ \citenamefont
  {Steimle}}]{Namiki1998}%
  \BibitemOpen
  \bibfield  {author} {\bibinfo {author} {\bibfnamefont {K.}~\bibnamefont {ichi
  C.~Namiki}}, \bibinfo {author} {\bibfnamefont {J.~S.}\ \bibnamefont
  {Robinson}},\ and\ \bibinfo {author} {\bibfnamefont {T.~C.}\ \bibnamefont
  {Steimle}},\ }\bibfield  {title} {\bibinfo {title} {A spectroscopic study of
  caoch$_3$ using the pump/probe microwave and the molecular beam/optical stark
  techniques},\ }\href {https://doi.org/10.1063/1.477146} {\bibfield  {journal}
  {\bibinfo  {journal} {The Journal of Chemical Physics}\ }\textbf {\bibinfo
  {volume} {109}},\ \bibinfo {pages} {5283} (\bibinfo {year}
  {1998})}\BibitemShut {NoStop}%
\bibitem [{\citenamefont {Petrov}\ and\ \citenamefont
  {Zakharova}(2022)}]{petrov2022sensitivity}%
  \BibitemOpen
  \bibfield  {author} {\bibinfo {author} {\bibfnamefont {A.}~\bibnamefont
  {Petrov}}\ and\ \bibinfo {author} {\bibfnamefont {A.}~\bibnamefont
  {Zakharova}},\ }\bibfield  {title} {\bibinfo {title} {Sensitivity of the yboh
  molecule to p t-odd effects in an external electric field},\ }\href@noop {}
  {\bibfield  {journal} {\bibinfo  {journal} {Physical Review A}\ }\textbf
  {\bibinfo {volume} {105}},\ \bibinfo {pages} {L050801} (\bibinfo {year}
  {2022})}\BibitemShut {NoStop}%
\bibitem [{\citenamefont {Ilia{\v{s}}}\ and\ \citenamefont
  {Saue}(2007)}]{iliavs2007infinite}%
  \BibitemOpen
  \bibfield  {author} {\bibinfo {author} {\bibfnamefont {M.}~\bibnamefont
  {Ilia{\v{s}}}}\ and\ \bibinfo {author} {\bibfnamefont {T.}~\bibnamefont
  {Saue}},\ }\bibfield  {title} {\bibinfo {title} {An infinite-order
  two-component relativistic hamiltonian by a simple one-step transformation},\
  }\href@noop {} {\bibfield  {journal} {\bibinfo  {journal} {The Journal of
  Chemical Physics}\ }\textbf {\bibinfo {volume} {126}},\ \bibinfo {pages}
  {064102} (\bibinfo {year} {2007})}\BibitemShut {NoStop}%
\bibitem [{\citenamefont {Saue}(2011)}]{saue2011relativistic}%
  \BibitemOpen
  \bibfield  {author} {\bibinfo {author} {\bibfnamefont {T.}~\bibnamefont
  {Saue}},\ }\bibfield  {title} {\bibinfo {title} {Relativistic hamiltonians
  for chemistry: A primer},\ }\href@noop {} {\bibfield  {journal} {\bibinfo
  {journal} {ChemPhysChem}\ }\textbf {\bibinfo {volume} {12}},\ \bibinfo
  {pages} {3077} (\bibinfo {year} {2011})}\BibitemShut {NoStop}%
\bibitem [{\citenamefont {Zou}\ \emph {et~al.}(2011)\citenamefont {Zou},
  \citenamefont {Filatov},\ and\ \citenamefont {Cremer}}]{zou2011development}%
  \BibitemOpen
  \bibfield  {author} {\bibinfo {author} {\bibfnamefont {W.}~\bibnamefont
  {Zou}}, \bibinfo {author} {\bibfnamefont {M.}~\bibnamefont {Filatov}},\ and\
  \bibinfo {author} {\bibfnamefont {D.}~\bibnamefont {Cremer}},\ }\bibfield
  {title} {\bibinfo {title} {Development and application of the analytical
  energy gradient for the normalized elimination of the small component
  method},\ }\href@noop {} {\bibfield  {journal} {\bibinfo  {journal} {The
  Journal of Chemical Physics}\ }\textbf {\bibinfo {volume} {134}},\ \bibinfo
  {pages} {244117} (\bibinfo {year} {2011})}\BibitemShut {NoStop}%
\end{thebibliography}
\end{document}